\documentclass[journal=jctcce,manuscript=article]{achemso}
\setkeys{acs}{maxauthors=0,articletitle=true}
\usepackage{achemso}
\usepackage{placeins}
\usepackage{graphics}
\usepackage{amssymb,amsfonts}
\usepackage{graphicx}
\usepackage[table,dvipsnames]{xcolor}
\usepackage{multirow}
\usepackage[font=footnotesize]{caption}
\usepackage{subcaption}
\usepackage{booktabs}
\usepackage{colortbl}
\usepackage{amsmath}
\usepackage{amsopn}
\usepackage{bm}
\usepackage{siunitx}
\usepackage{bm}
\usepackage{color}
\usepackage{array}
\usepackage{lscape}
\usepackage{mciteplus}
\usepackage[version=3]{mhchem}
\usepackage{ulem}
\usepackage{listings}
\usepackage{enumerate}
\usepackage{lmodern}
\usepackage{scalerel,stackengine}
\usepackage{tabularx}
\usepackage{longtable}

\DeclareMathOperator{\tr}{Tr}
\DeclareMathOperator\erfc{erfc}

\author{Luna Zamok}
\affiliation{DTU Chemistry, Technical University of Denmark\\Kemitorvet Bldg. 206, 2800 Kgs. Lyngby, Denmark}
\author{Janus J. Eriksen}
\email{janus@dtu.dk}
\affiliation{DTU Chemistry, Technical University of Denmark\\Kemitorvet Bldg. 206, 2800 Kgs. Lyngby, Denmark}

\title[TITLE]{Atomic Decompositions of Periodic Electronic-Structure Simulations}

\begin{document}
    
\begin{abstract}

We present a new theory for partitioning simulations of periodic and solid-state systems into physically sound atomic contributions at the level of Kohn-Sham density functional theory. Our theory is based on spatially localized linear combinations of crystalline Gaussian-type orbitals and, as such, capable of exposing local features within periodic electronic structures in a more intuitive and robust manner than alternatives based on the spatial distribution of atomic basis functions alone. Early decomposed cohesive energies of both molecular polymers and different crystalline polymorphs demonstrate how the atomic properties yielded by our theory convincingly align with the expected charge polarization in these systems, also whenever partial charges and Madelung energies may lend themselves somewhat ambiguous to interpretation.

\end{abstract}

\newpage

\section{Introduction}\label{intro_sect}

While the majority of all solids are repeating at extended length scales, most of the electronic properties of key interest derived from such systems will be intrinsically local in nature. Thus, instead of attempting to simulate their physics at large for a truly insurmountable particle number, one typically opts to describe crystals by identifying a suitably isolated cluster---itself saturated in all outwards directions---or, as is preferable, by introducing periodic boundary conditions (PBCs) around a primitive cell object. In the latter case, Bloch's theorem will allow for summations over an infinite number of translations to be replaced by integrals over crystal momentum vectors ($\bm{k}$) inside the first Brillouin zone, integrals which may themselves, in practice, be appropriately approximated as weighted sums over discrete points in reciprocal space. For example, the standard Monkhorst-Pack algorithm formally expresses such integrals through an equidistant grid of $\bm{k}$-vectors with identical weights~\cite{Monkhorst-Pack_kmesh}.\\

In comparison with routine chemical simulations of molecules in isolation, it may be convenient to view the above transformation from real to reciprocal space for systems composed of repeating units as involving an effective downfolding from an infinite system to a lattice. Such a preliminary step is naturally not part of standard simulations of chemical Hamiltonians. Unit cells thus represent the basic subunits from which to construct larger cells, but what now if one desires to decompose periodic simulations even further, e.g., into properties associated with the individual atoms of said cell? For instance, the ordered arrangement of atoms in a crystalline state is typically accompanied by a net gain in internal energy, but how may one probe contributions on a truly local scale to such global, cohesive properties? Will atomic properties align with expectation and intuition based on prior knowledge of charge polarization in a given solid and the oxidation states of its atoms, and what will the sensitivity of predictions like these amount to under a change of simulation parameters?\\

For similar studies within organic chemistry, we have recently proposed a novel means of partitioning first-order properties of molecules in either ground or excited electronic states at the mean-field level, i.e., Hartree-Fock or Kohn-Sham density functional theory (KS-DFT)~\cite{eriksen_decodense_jcp_2020,eriksen_elec_ex_decomp_jcp_2022}. By rotating the canonical orbitals of standard electronic-structure simulations into a spatially localized representation, we have demonstrated how our decompositions may allow for probing local, atomic properties within condensed (bulk-like) phases~\cite{eriksen_local_condensed_phase_jpcl_2021}, as well as the study of emerging trends among contributions to atomization energies from distinct elements embedded within molecules. For the purpose of training high-dimensional neural networks to predict both atomic and molecular properties~\cite{eriksen_nn_qm7_decomp_jctc_2023}, but also in the physicochemical interpretation of local electronic structures~\cite{eriksen_atom_energy_jctc_2023}, we have found our proposed decomposition scheme---expressed in a localized basis of molecular orbitals (MOs)---to be arguably more robust, consistent, and physically sound than alternatives that instead operate exclusively on the basis of how the underlying atomic orbitals (AOs) are distributed within a molecule.\\

For molecular systems, MOs are almost always expanded in a set of basis functions consisting of Gaussian-type orbitals (GTOs)~\cite{mest}. However, under the leading assumption that the presence of atoms in a crystal can be treated as mere slight distortions to a free-electron picture, simulations of solids tend to use bases of plane waves instead~\cite{martin_elec_struct}. While these are independent of nuclear positions, mutually orthogonal, and exhibit no basis set superposition errors, a great number of plane waves are typically required for achieving a sufficient level of accuracy in solid-state simulations. In addition, the non-local nature of this approach renders most chemical information lost due to the lack of atomic quantities. Embedding formalisms have thus been proposed as a way to recover locality~\cite{goodpaster_embedding_jctc_2018}, as have unitary transformations of delocalized wave functions~\cite{soler_projection_sscomm_1995,dronskowski_jcc_2013}, e.g., by means of analytical projection-based techniques in the popular {\texttt{LOBSTER}} code~\cite{lobster_jcc_2016,lobster_jcc_2020}. Further to that, Wannier functions can be used as an effective representation of Bloch states~\cite{vanderbilt_wannier_rmp_2012}, as is true also for more recent frozen-density embedding methods~\cite{dronskowski_lmo_jpca_2023}. Alternatively, one may opt for crystalline GTOs as the basis functions for a periodic simulation of solid-state materials and proceed through a machinery much reminiscent of that employed in standard simulations of chemical Hamiltonians, namely, by constructing MOs as linear combinations of AOs adapted with respect to translational symmetry.\\

To probe local, atomic features within periodic solids, standard charge populations, derived from either Hilbert or real-space protocols, have long been used to detect ionic crystalline bonds. Atomic partial charges also serve as precursors in the computation of Madelung energies, the Coulombic part of lattice energies, and thus supposed indicators of electrostatic stabilization in many crystal structures~\cite{kageyama_jacs_2017,dronskowski_acie_2021}. In the case of an ionic crystal, the Madelung energy obviously makes up a major constituent of the total cohesive energy, whereas in simulations of phases in which absolute electronegativities hardly differ at all, agreements with formation enthalpies (e.g., total energies at pressures close to zero) may often tend not to materialize. As an alternative for examining crystalline structures, covalent, intermediate, and ionic bonding interactions have been studied through different proxies to determine interatomic bond strengths~\cite{hughbanks_hoffmann_jacs_1983,dronskowski_blochl_jpc_1993,dronskowski_jpca_2011,dronskowski_jpcc_2021,george_hautier_cpc_2022}, e.g., for detecting important interactions, mapping out entire bond graphs, and determining coordination environments. These types of studies, however, often require matching comparisons to purely geometric approaches, just like analyses based on atomic charges need to be correlated with {\textit{a priori}} assumed oxidation states.\\

In the present work, we instead propose a new scheme for decomposing total energies (at no associated loss) into atomic contributions without introducing any biases toward either structural or electronic effects in solids. We will study whether or not these local contributions provide insights into periodic structures, akin to the observations made in a recent study of molecular systems~\cite{eriksen_atom_energy_jctc_2023}, where atomic contributions showed promise as distinct fingerprints of atoms embedded within molecules. We will also study how atomic contributions to cohesive energies from different AO- and MO-based partitioning schemes align with intuition, for instance, when comparing relative stabilities of different crystalline polymorphs.\\

Recently, the authors of Ref. \citenum{dronskowski_rsc_adv_2019} proposed an effective way of performing both Mulliken- and L{\"o}wdin-type population analyses for solids through a combination of both projector augmented-waves and a local AO basis set within the {\texttt{LOBSTER}} code. Importantly, these charges were demonstrated not to exhibit the same strong dependency on basis set as is known from simulations on molecules, remaining reasonably stable with respect to an increase in the kinetic energy of the plane waves and, as such, avoiding an otherwise costly partitioning of plane-wave electron densities in real space. In practice, Ref. \citenum{dronskowski_rsc_adv_2019} proposed to project crystal wave functions onto an auxiliary basis of contracted Slater-type orbitals, upon which atomic populations are determined. In here, we will demonstrate the importance of a similar type of operation, namely, one in which crystalline MOs are projected from a full, computational basis onto a minimal, free-atom basis of intrinsic AOs (IAOs)~\cite{knizia_iao_ibo_jctc_2013,janowski_quambo_jctc_2014}. Despite having been tailored towards molecular applications (in the absence of pseudopotentials), we will demonstrate how atomic populations based on IAOs render Mulliken-like charges that are significantly less sensitive to the composition of the computational basis than standard charges, on balance with similar observations made for molecular applications in Ref. \citenum{lehtola_jonsson_pm_jctc_2014}.\\

The present study is outlined as follows. In Sect.~\ref{theory_sect}, we discuss what additions to a standard implementation of periodic KS-DFT are required to partition simulations of total energies among the individual atoms of a system by means of either crystalline AOs or MOs. Next, in Sect.~\ref{comp_details_sect}, we comment on the computational details of our numerical studies to follow in Sect.~\ref{results_sect}, where results are presented for both several small molecules and molecular polymers as well as distinct polymorphs of two different solids. Finally, a brief summary and some concluding remarks on the potential of our new theory are presented in Sect.~\ref{summary_concl_sect}.

\section{Theory}\label{theory_sect}

As mentioned above, the total KS-DFT energy in the presence of PBCs is computed by sampling either successively larger supercells at the $\Gamma$-point or, alternatively, by integrating the first Brillouin zone of the reciprocal unit cell using an appropriate mesh of $\bm{k}$-points. The former of these two approaches is computationally more cumbersome, as it does not allow for explicitly enforcing translational symmetry and conserving crystal momentum, but the two are, however, equivalent whenever a supercell comprises an arrangement of unit cells in accordance with a sampled wave vector. For a periodic Hamiltonian formulated in a basis of crystalline GTOs, the KS-DFT energy can---on par with chemical Hamiltonians---be expressed as a functional of the one-body reduced density matrix (1-RDM), $\bm{D}(\bm{k})$, in a sum over all wave vectors (of which we have $N_k$ in total) within the first Brillouin zone:
\begin{align}
    E_{\text{KS-DFT}} 
    = \frac{1}{N_k}\sum_{\bm{k}}
    &(\tr[(\bm{T}_{\text{kin}}(\bm{k}) + \bm{V}_{\text{nuc}}(\bm{k}))\bm{D}(\bm{k})] 
    \nonumber 
    \\
    &+ \frac{1}{2}\sum_{\sigma}\tr[\bm{V}_{\text{eff},\sigma}(\bm{D})\bm{D}_{\sigma}(\bm{k})] + 
    \tr[\epsilon_{xc}(\bm{\rho})\bm{\rho}(\bm{k})]) + E_{\text{struct}} \ .
    \label{eq-e_dft_total}
\end{align}
In Eq. \ref{eq-e_dft_total}, the $\sigma$-subscript refers to spin, while $\bm{T}_{\text{kin}}(\bm{k})$, $\bm{V}_{\text{nuc}}(\bm{k})$, and $\bm{V}_{\text{eff}}(\bm{D})$ denote the kinetic energy operator, nuclear attraction operator, and effective Fock potential, respectively. The latter depends on the total 1-RDM, $\bm{D}$, which is a (weighted) sum of $\{{\bm{D}}({\bm{k}})\}$. The exchange-correlation ($xc$) energy is expressed in terms of the computed energy density, $\epsilon_{xc}(\bm{\rho})$, as derived from the total electronic density, $\bm{\rho}$, and possibly its derivatives, all quantities which may be defined by proceeding through the total 1-RDM, $\bm{D}$. Finally, $E_{\text{struct}}$ is the scalar structural repulsion energy. Alternatively, in supercell simulations, rather than computing the total KS-DFT energy of a unit cell as a weighted sum across a mesh of $\bm{k}$-vectors, computational formula are used akin to those valid for molecular systems, but using operators expressed in a GTO basis set over the extended reciprocal lattice.\\ 

In systematically decomposing total KS-DFT energies for periodic systems in a way that maps individual contributions onto cell structures, we will, first and foremost, require an unbiased treatment of structural and electronic contributions. An obvious way of achieving such a partitioning among atoms is thus to restrict the trace operations in Eq. \ref{eq-e_dft_total} to run over only those AOs that are spatially centered on specific nuclei. This way of partitioning the 1-RDMs will give rise to a scheme reminiscent of one by Nakai for molecular Hamiltonians, denoted as the energy density analysis (EDA)~\cite{nakai_eda_partitioning_cpl_2002,nakai_eda_partitioning_ijqc_2009}. Straightforward extensions of the EDA scheme to solids are thus trivial whenever formulated in a basis of GTOs, although the strong dependence on the composition of a given one-electron basis set known from quantum-chemical applications will likely prevail and thereby render it excessively sensitive to the full computational AO basis of choice. Furthermore, as has recently been demonstrated for a selection of prototypical organic species~\cite{eriksen_atom_energy_jctc_2023}, the EDA scheme (alongside its possible extensions~\cite{baba2006natural,imamura2007grid}) is bound to yield atomic contributions that are largely predetermined and, hence, insensitive to variations in the underlying electronic structure. As part of the present work, we will study to what extent this conclusion holds true for periodic Hamiltonians as well. More details on our implementations are provided in the supporting information (SI).\\

Moving beyond schemes tied to the AO basis, we have recently proposed how to exploit the fact that a total 1-RDM can be decomposed into a sum of contributions from individual MOs, which, in turn, may be grouped on the basis of which atoms of a system these populate~\cite{eriksen_decodense_jcp_2020}. Given how the total energy stays invariant under a unitary MO transformation, we have previously demonstrated how suitable choices of spatially localized MOs and their atomic populations can return a set of atom-specific 1-RDMs, $\{\bm{\delta}\}$, to be traced with the operators in Eq. \ref{eq-e_dft_total}, which yield physically sound atomic contributions to total energies. We will here proceed to extend the theory in Ref. \citenum{eriksen_decodense_jcp_2020} to the study of periodic solids as well, albeit only for systems with a finite band gap (in the absence of smearing) in its current form. \\

As MO coefficients are complex in reciprocal space, we generally begin by transforming (by means of an inverse fast Fourier procedure) converged MOs and electron repulsion integrals to corresponding quantities defined for a supercell object at the $\Gamma$-point before decomposing energies on the basis of AOs or MOs. Following these transformations, evaluations of the kinetic and $xc$ energy terms for a periodic system are formally analogous to the case of a chemical Hamiltonian, i.e., these are traced by either total or atom-specific 1-RDMs.\\

For the sake of computational efficacy, we may opt to replace core electrons by norm-conserving pseudopotentials (PPs) in our KS-DFT simulations, namely, the separable dual-space Gaussian variants developed by Goedecker, Teter, and Hutter\cite{GTH_pp,GTH_pp_rel} (GTH), as implemented in {\texttt{PySCF}}~\cite{berkelbach_eom_cc_jctc_2017,pyscf_wires_2018,pyscf_jcp_2020}. These potentials, which effectively replace the standard nuclear attraction operator, $\bm{V}_{\text{nuc}}$, in Eq. \ref{eq-e_dft_total}, consist of both local and non-local components, collectively intended to remove sharp nuclear densities. Given how the total operator, $\bm{V}^{\text{GTH}}$, is defined as a sum over ion-specific contributions in either real or reciprocal space, the handling of these in any partitioning scheme will not differ much from the case of all-electron simulations where $\bm{V}_{\text{nuc}} = \sum_K\bm{V}_{K}$. 
However, as recently discussed in Ref. \citenum{eriksen_atom_energy_jctc_2023}, to distinguish, as a frame of reference, between ({\textit{i}}) an atom and ({\textit{ii}}) its associated electrons and the interactions between these and all other $\mathcal{M}-1$ nuclei and their assigned electrons, respectively, we divide the contraction between $\bm{V}^{\text{GTH}} = \sum_K\bm{V}^{\text{GTH}}_K$ and $\bm{D}$ on par with earlier work~\cite{eriksen_decodense_jcp_2020}:
\begin{align}
    \tr[\bm{V}^{\text{GTH}}\bm{D}] 
    = \frac{1}{2} \sum_K(\tr[\bm{V}^{\text{GTH}}_{K}\bm{D}] + \tr[
    \bm{V}^{\text{GTH}}\bm{\delta}_{K}])
    \label{gth-eq}
\end{align}
where $\{\bm{\delta}\}$ are the atom-specific 1-RDMs discussed above (cf. the SI). In the EDA scheme, the first term remains the same whereas the second term is once more replaced by a truncated trace over only those AOs ($\mu$) that are spatially assigned to atom $K$, i.e., $\tr_{\mu\in K}[\bm{V}^{\text{GTH}}\bm{D}]$. Collectively, the first term of Eq. \ref{gth-eq} (alongside the analogous contribution from $\bm{V}_{\text{nuc}}$) will be denoted as $E^{\text{(1)}}_{\text{ne}}$, while the latter contributions from $\bm{V}^{\text{GTH}}$ and $\bm{V}_{\text{nuc}}$ will be denoted as $E^{\text{(2)}}_{\text{ne}}$.\\

Unlike when simulating isolated chemical systems, the handling of long-range Coulombic interactions will further differ in that these need to be evaluated for an infinite system in the presence of periodic boundaries. In turn, this changes the way one computes the (scalar) repulsion between nuclei, $E_{\text{struct}}$, the effective Fock potential, $\bm{V}_{\text{eff}}$, and the attraction between nuclei and electrons arising from a periodic potential, $\bm{V}_{\text{nuc}}$ (or $\bm{V}^{\text{GTH}}$). Each of these terms will be divergent on their own, in that one needs to explicitly include long-range electrostatics between the nuclei (and electrons) of a periodic lattice, resulting in infinite sums over these interactions. To remedy this for the electronic interactions, we exclude the contribution from the reciprocal lattice vector, $\bm{G}=0$, in the effective Fock potential as well as in the periodic potentials, $\bm{V}_{\text{nuc}}$ and $\bm{V}^{\text{GTH}}$. In the latter case, we remove these contributions stemming from each specific ion separately. Furthermore, whenever $\bm{V}^{\text{GTH}}$ is used for treating nuclear-electron interactions, the functional form of the short-range part deviates from that of a plain Coulomb potential, and, as a result, an additional term consisting of a summation over the ionic cores in the unit cell will need to be evaluated~\cite{berkelbach_eom_cc_jctc_2017}. Overall, this renders a partitioning of electronic electrostatics into atomic contributions feasible, although the computational implementations of these terms in {\texttt{PySCF}} differ slightly depending on which density-fitting scheme is used, based on GTOs~\cite{GDF_MDF_PySCF} (GDF) or plane waves~\cite{CP2K_GPW_also_pyscf,CP2K} (FFTDF).\\

As for the Ewald treatment of short- and long-range nuclear repulsion, these are divided into convergent real ($E^{\bm{R}}$) and reciprocal ($E^{\bm{G}}$) summations, both of which may be conveniently expressed in terms of contributions inherent to each of the atoms of a unit cell~\cite{berkelbach_eom_cc_jctc_2017,ewald_sum}:
\begin{subequations}
\begin{align}
    E^{\bm{R}}_{K} 
    &= 
    \frac{Z_K}{2}
    \Big[
    \sum_{L}
    Z_L
    \sum_{\bm{T}}
    \frac{\erfc{(\eta|\bm{R}_K - \bm{R}_L - \bm{T}|)}}
    {|\bm{R}_K - \bm{R}_L - \bm{T}|}
    -
    \frac{2\eta}{\sqrt{\pi}}
    Z_K
    -
    \frac{\pi}{\eta^2\Omega}
    \sum_K
    Z_K 
    \Big] 
    \label{eq-ewald_T} \\
    E^{\bm{G}}_{K} 
    &= 
    \frac{Z_K}{2}
    \Big[
    \frac{4\pi}{\Omega}
    \sum_{\bm{G}\neq\bm{0}}
    S^{\bm{G}}_K
    (
    \sum_{K}
    Z_K S^{\bm{G}}_K
    )^{\ast}
    \frac{\exp{(-G^2/4\eta^2)}}{G^2} 
    \Big] \ 
    \label{eq-ewald_G}
\end{align}
\end{subequations}
In Eq. \ref{eq-ewald_T}, the self-interaction terms ($K = L$) in the summation over lattice translation vectors, $\bm{T}$, are implicitly neglected whenever $\bm{T} = \bm{0}$. The structural factors in Eq. \ref{eq-ewald_G} are defined as $S^{\bm{G}}_K = \exp{(i\bm{G}\bm{R}_K)}$, and the volume of the unit cell is denoted as $\Omega$. As per convention, both the range of the lattice summations as well as the Ewald splitting parameter, $\eta$, are chosen so as to facilitate rapid convergence of the total structural contribution.

\section{Computational Details}\label{comp_details_sect}

In generating the results to follow in Sect.~\ref{results_sect}, the PBE~\cite{perdew_burke_ernzerhof_pbe_functional_prl_1996} $xc$ functional was used in the simulations of H$_2$O, NH$_3$, CH$_4$ GaH$_3$, and TiCl$_4$ in Sect.~\ref{fingerprints_subsect} as well as GaN in Sect. \ref{gan_subsect}, while the BLYP~\cite{B_lyp,b_L_yp,bl_YP} and PBEsol~\cite{PBEsol1,PBEsol2} functionals were used in Sects.~\ref{fingerprints_subsect} and \ref{batio3_subsect} in the simulations of the PAN polymer and BaTiO$_3$, respectively. All $xc$ functionals were accessed through {\texttt{Libxc}}~\cite{libxc_software_x_2018}. In the absence of PBCs, standard all-electron DZVP basis sets were used for the PAN polymer~\cite{DZVP}, while all-electron and GTH optimized basis sets of double-$\zeta$ quality were used in the periodic simulations~\cite{gth_molopt_basis}. The latter were used in all calculations that employed GTH pseudopotentials, while all other all-electron calculations used the Pople-style 6-31G* basis set~\cite{631Gs_H_ditchfield1971a,631Gs_CNO_hehre1972a,631Gs_CNO_hariharan1973a,631Gs_Ti_rassolov1998a,631Gs_Ga_rassolov2001a}. In all instances, electron-repulsion integrals were computed in {\texttt{PySCF}} using Gaussian density fitting with Weigend auxiliary basis sets~\cite{weigend_aux_basis}, and linear AO dependencies were eliminated through an initial Cholesky orthogonalization, as proposed by Lehtola~\cite{lehtola_cholesky_orth_jcp_2019}.\\

One-dimensional grids of between 1 to 7 $\bm{k}$-points were used to compute KS-DFT energies for the PAN polymer, while three-dimensional, equidistant grids of between 1 to $3^3$ $\bm{k}$-points were used in the simulations of GaN and BaTiO$_3$. 
MOs were next transformed to corresponding supercells, before being rotated into a basis of either standard Pipek-Mezey~\cite{pipek_mezey_jcp_1989} (PM) MOs or intrinsic bond orbitals (IBOs), for which the localization operates on IAO-based populations~\bibnote{In constructing IAOs, the MINAO basis was used for all elements but Ba, for which the ANO-RCC-MB minimal basis was used instead.}. In the MO-based partitioning scheme of Sect.~\ref{theory_sect}, either standard or IAO-based Mulliken populations were subsequently used as weights to partition total energies using the \texttt{decodense} code~\cite{decodense}, which was also used for the orbital-invariant EDA partitionings.

\section{Results}\label{results_sect}

As in earlier works of ours~\cite{eriksen_atom_energy_jctc_2023}, all results will be presented as cohesive energies, i.e., differences between decomposed and free-atom energies (calculated in unit cells of volume dimensions $\Omega = (30 \ \text{\AA})^3$). Additional results are reported in the supporting information (SI).

\subsection{Validation}\label{fingerprints_subsect}

\begin{figure}[htbp]
    \centering
    \includegraphics[width=\textwidth]{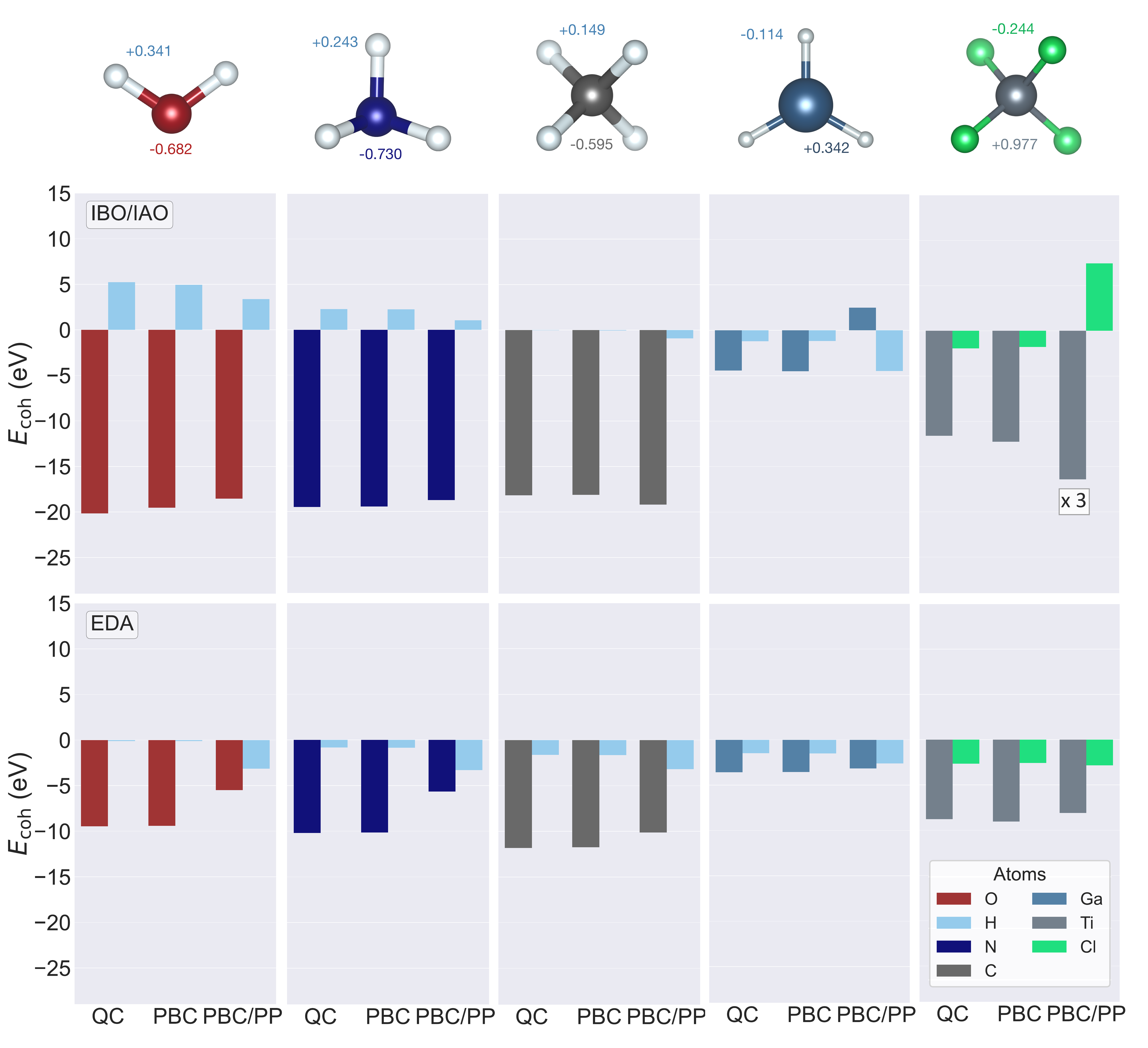}
    \caption{Atomic cohesive energies for H$_2$O, NH$_3$, CH$_4$, GaH$_3$, and TiCl$_4$ computed via standard quantum-chemical (QC) simulations or using PBCs, either without or with pseudopotentials and corresponding basis sets (PBC/PP). The insets show IAO-based Mulliken charges derived from all-electron PBC simulations.}
    \label{fig_1}
\end{figure}
We start by validating our periodic implementations of both the AO- (EDA) and MO-based (IBO/IAO) partitioning schemes of Sect. \ref{theory_sect}. To allow for comparisons to be made to the organic chemistry in Ref. \citenum{eriksen_atom_energy_jctc_2023}, we compute atomic cohesive energies for three simple molecules---H$_2$O, NH$_3$, and CH$_4$---alongside two inorganic compounds involving heavier atoms from the fourth period of the table of elements---GaH$_3$ and TiCl$_4$---using both standard quantum-chemical (QC) simulations, equivalent simulations in the presence of PBCs, as well as periodic simulations with core regions effectively replaced by GTH pseudopotentials (PBC/PP), the latter two both simulated at the $\Gamma$-point. In the former QC case, molecules are naturally simulated in vacuum, whereas in the latter two periodic cases, they were all confined within deliberately large cells ($\Omega = (10 \ \text{\AA})^3$) to artificially reduce replicating effects. A convergence test showing how this volume is indeed sufficient is provided in Fig. S1 of the SI.\\

Fig.~\ref{fig_1} depicts cohesive energies of all constituent atoms decomposed using either the EDA or IBO/IAO schemes (see also Table S1 of the SI). Starting with the organic compounds, the contributions associated with H, C, N, and O atoms are consistently observed to be similar to the reference values computed for the alkanes, amines, and oxygen-containing compounds in Ref. \citenum{eriksen_atom_energy_jctc_2023}, regardless of the exact manner by which the underlying simulations are performed. However, while total results for the QC, PBC, and PBC/PP simulations reported here are very much on par for H$_2$O, NH$_3$, and CH$_4$, Fig. S2 of the SI reveals how individual contributions to the atomic cohesive energies may differ substantially (sometimes more than doubling in magnitude). The MO-based IBO/IAO scheme yields a net stabilization of C, N, and O atoms, and thus a corresponding minor destabilization of all hydrogens with respect to their state in the gas phase. In contrast, all atoms are predicted to be stabilized in the results of the AO-based EDA scheme, again rendering their effective fingerprints indistinguishable. Also, as discussed in Ref. \citenum{eriksen_atom_energy_jctc_2023}, the EDA energies are bound to change substantially upon augmenting the basis set with diffuse functions, unlike those of the MO-based IBO/IAO scheme.\\ 

For the two inorganic compounds, however, the picture is observed to change somewhat. Here, the atomic cohesive energies of the IBO/IAO scheme are all negative in both sets of all-electron calculations, resembling the results obtained in the EDA scheme rather closely. Again, the introduction of PBCs induces only minor shifts to the atomic cohesive energies, whereas the use of PPs is now observed to shift all energies more distinctively, particularly those of the IBO/IAO scheme, cf. Table S1 of the SI. For TiCl$_4$, the atomic cohesive energy of Ti changes by almost 40 eV (as indicated in Fig. \ref{fig_1} by `x3' to ease comparison between panels), and the Cl atoms are predicted to be destabilized in the compound over their gas-phase reference state as a result. Likewise, in GaH$_3$, the cohesive energy of Ga changes sign in the presence of PPs. The EDA results, on the other hand, are mostly observed not to change significantly upon introducing PPs, on account of the strict dependence on the locality of individual AOs. To rule out the use of the standard MINAO basis for computing IAO-based populations as the cause of the dissimilar IBO/IAO-based results in Fig. \ref{fig_1}, Fig. S4 of the SI shows how analogous results are obtained using a GTH-adapted minimal basis instead.\\

In this context, it is informative to recall how many atomic quantities are known to change substantially whenever a PP is used in lieu of an explicit treatment of the core region of heavier atoms. As elaborated upon in Ref. \citenum{coh_en_PP_goedecker}, the spin polarization of an isolated atom is inadequately described when invoking PPs, a formal deficiency which is thus bound to propagate into the computation of atomic cohesive energies. Fig. S3 of the SI examines the individual contributions to the energies in Fig. \ref{fig_1}, further revealing how these change by several orders of magnitude for Ga and Ti atoms upon introducing PPs, but also how the final shifts to total results stay systematic. Generally speaking, this is not an issue of any practical importance per se whenever relative rather than absolute atomic quantities are desired, given how errors will eventually cancel out. For instance, differences in atomic cohesive energies between solid states may be expected to be comparable with or without PPs ({\textit{vide infra}}), as would changes to atomic energy contributions along a reaction coordinate.\\

As touched upon in Sect.~\ref{intro_sect}, two approaches can be adopted when evaluating properties for extended systems; the first of these involves computing the desired property using PBCs, while in the latter one represents the bulk system by an adequately sized cluster. The conceptual differences between these two approaches can be easily illustrated. In the case of a periodic polymer along a single dimension, the use of PBCs can be conceptualized as representing the infinite polymer as a closed ring object. Each equivalent atom on this ring will experience the same local environment, and so a physical decomposition scheme should be expected to yield a set of partitioned energies that are uniform. In contrast, if one is to simulate the same local property by means of conventional quantum chemistry, the polymer must be modelled as an elongated chain and the corresponding local atomic environment within this finite chain will be most accurately represented from its most central segment.\\

\begin{figure}[htbp]
    \centering
    \includegraphics[width=0.9\textwidth]{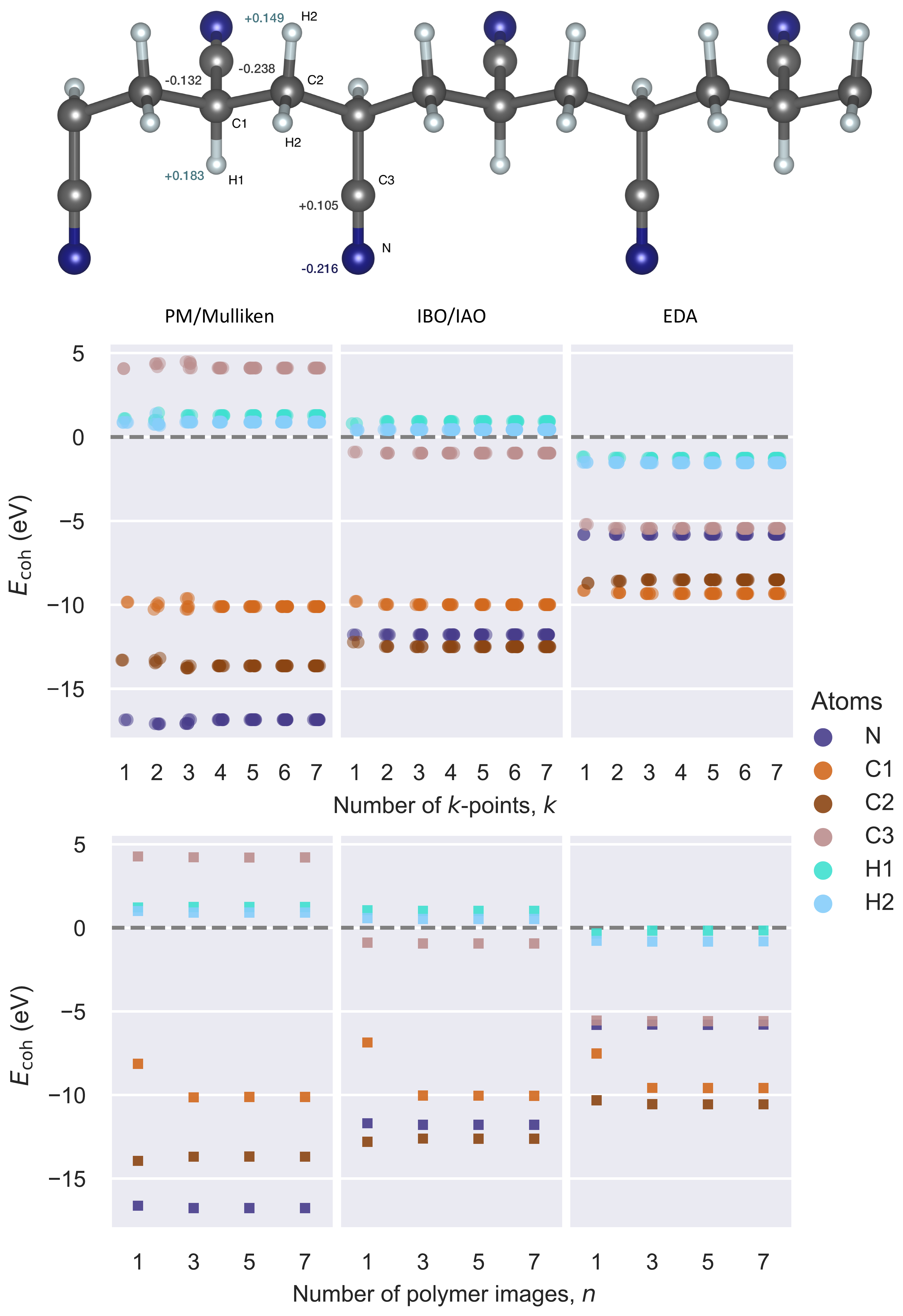}
    \caption{Atomic cohesive energies for a PAN polymer, computed using all-electron basis sets with PBCs (upper panel) or without (lower panel). The inset shows IAO-based Mulliken charges simulated using PBCs.}
    \label{fig_2}
\end{figure}
To further gauge our PBC implementations of the AO- and MO-based partitioning schemes of Sect.~\ref{theory_sect}, we next decompose KS-DFT energies of a simple organic polymer, polyacrylonitrile (PAN), cf. Fig.~\ref{fig_2}, using either PBCs (with $\bm{k}$-point sampling along the periodicity direction) or a standard quantum-chemical approach (with an increasing number of repeating PAN polymer units). The cohesive energy contributions depicted are those from the nitrogen, the three distinct carbons atoms, and the two hydrogens. To complement our previous results, we additionally decompose energies using an alternative MO-based scheme, namely, one that relies on PM orbitals and corresponding standard Mulliken populations.\\

By contrasting the periodic and molecular simulations in Fig.~\ref{fig_2}, all results of the former are observed to converge significantly faster with problem size than those of the latter, showing only minor variations upon moving beyond the $\Gamma$-point. In comparing the AO- and MO-based partitioning schemes, the first noticeable difference between the two is that the former again predicts a net stabilization with respect to the gas phase for all atoms. This is true regardless of whether or not the simulations are subject to PBCs. Both MO-based partitionings, on the other hand, point to net stabilizations of only the (non-cyano) C and N atoms, a pattern much reminiscent of what was observed above for the isolated molecules. In particular, the IBO/IAO energies are found to match those for the smaller acetonitrile system reported in Ref. \citenum{eriksen_atom_energy_jctc_2023} convincingly well. However, the contribution from the cyano carbon (C3) in the PM/Mulliken scheme marks a noticeable exception to this general trend, predicting a considerable destabilization of this carbon atom in both types of simulations.\\

Although different basis sets are used in the (non-)periodic simulations, the cohesive energies are very much on a comparable scale between the MO-based partitionings, in particular, whenever a combination of IBOs and IAO weights are used in these. The same holds true even whenever core electrons are represented by PPs in the PBC simulations, cf. Fig. S5 of the SI, although the aforementioned shift in energies is observed to affect some atoms slightly more than others. Finally, the similarity between PBC and standard QC simulations is further corroborated by comparing Mulliken partial charges derived from these, but, importantly so, only when these are based on an intermediate free-atom (IAO) basis, not when comparing standard charges computed in the full basis sets, cf. Table S2 of the SI.

\subsection{Gallium Nitride}\label{gan_subsect}

\begin{figure}[htbp]
    \centering
    \includegraphics[width=0.9\textwidth]{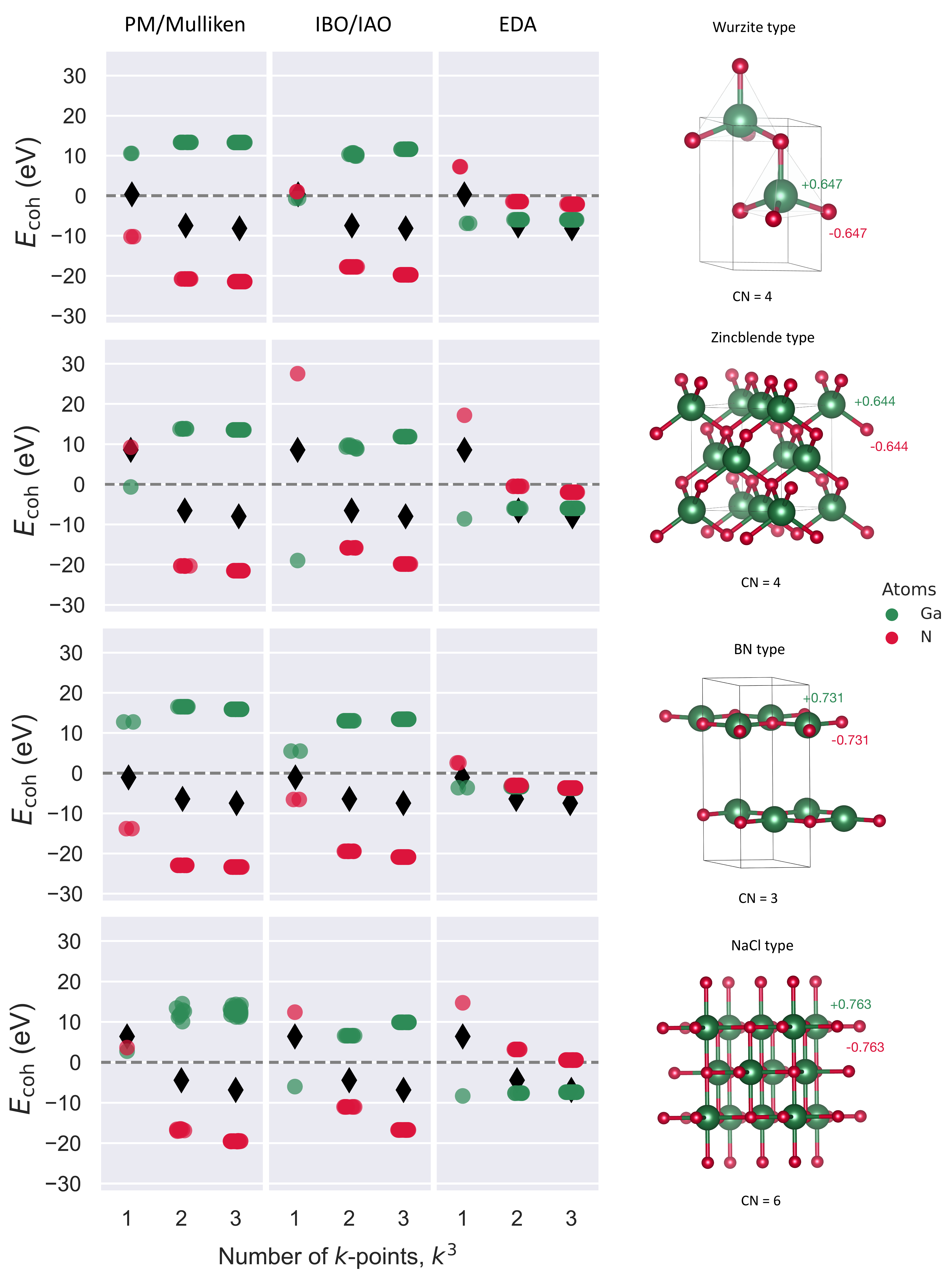}
    \caption{Atomic cohesive energies for different GaN polymorphs. Black markers represent total cohesive energies, and the panel on the right shows coordination numbers alongside IAO-based Mulliken charges.}
    \label{fig_3}
\end{figure}
In moving from an organic polymer periodic along just a single dimension to three-dimensional solids, we will start by studying gallium nitride (GaN), a well-known binary III/V direct bandgap semiconductor widely used in optoelectronic technologies~\cite{GaN_review}. In the present context, GaN is an interesting pilot example as one would expect its most stable phases to exhibit significant, yet somewhat diverse covalent interactions, as discussed in a recent bonding analysis by George and co-workers~\cite{george_hautier_cpc_2022}, even though relative phase stabilities are hard to rationalize on the basis of total cohesive energies alone as these are of the same size. Specifically, we have simulated the same four polymorphs of GaN as in Ref. \citenum{george_hautier_cpc_2022} (at 0 K), namely, the wurzite, zincblende, BN, and NaCl types~\bibnote{Structures have been loaded from the Materials Project database under the identifiers mp-804, mp-830, mp-1007824, and mp-2853.}, for which the coordination environment of Ga varies from tetrahedral in the former two to trigonal planar and octahedral in the latter two, respectively.\\

Our AO- and MO-based results using PPs are collected in Fig.~\ref{fig_3}. First, the convergence in $\bm{k}$-space is observed to be significantly slower than what was seen above for the organic polymer. However, although absolute atomic contributions are observed to vary some in moving from a grid of $2^3$ to $3^3$ $\bm{k}$-points, the relative ratios between these remain on par. The degree of polarization in each of the phases is thus the same, but the importance of moving beyond simulations only at the $\Gamma$-point is obvious across all different partitioning schemes. Next, traditional PM orbitals localized in the full computational basis are again observed to be inferior to IBOs. In particular, when a localization procedure is initiated from a set of canonical orbitals, using standard Mulliken populations, one is not guaranteed to arrive at a uniform set of energies for equivalent atoms, as is otherwise to be expected on account of symmetry arguments, cf. the PM/Mulliken results for the NaCl phase.\\

That being said, the two MO-based partitioning schemes in Fig.~\ref{fig_3} are seen to reveal much the same picture, i.e., a net stabilization of nitrogen over gallium across all four phases, whereas the AO-based EDA results observe an altogether opposite trend, with the scheme predicting a net stabilization of Ga. Recalling how nitrogen is significantly more electronegative than gallium, it would arguably seem fair to expect the shared electron density across all bonds to be localized closer to the nitrogens. Zooming in on the IBO/IAO results, the decrease in overall stability of the BN and NaCl phases, particularly the latter, appear to be related to changes in degrees of polarization. For instance, in the case of the NaCl phase, the nitrogens become less stabilized. This is likely due to an increase in steric repulsion between the nitrogens in this phase, a view which is further supported by the lowest predicted electrostatic stabilization for this symmetry, cf. Ref. \citenum{george_hautier_cpc_2022}, in which the authors also computed the highest character of covalent bonding in the phases with a tetrahedral arrangement of Ga atoms. The NaCl phase also features the longest bond distances and the largest differences in IAO-based partial charges, cf. Fig.~\ref{fig_3}. On the other hand, both IBO/IAO energies and IAO-based atomic charges within the wurzite and zincblende phases are virtually the same.\\

To complement the results in Fig.~\ref{fig_3}, but also the discussion on the use of PPs in Sect. \ref{fingerprints_subsect}, Fig. S6 of the SI presents corresponding all-electron results for the four phases of GaN. While the absolute energies are again observed to change with or without PPs, changes to the individual atomic cohesive energies of Ga and N atoms in-between different solid phases remain the same, which is testament to the expected cancellation of errors discussed above.\\

We will here refrain from discussing whether or not the bonds in the GaN phases are straightforwardly covalent, but we note that atomic decomposition schemes, particularly those based on the locality of MOs rather than AOs, are clearly capable of exposing subtle differences between otherwise very similar phases, thus allowing for local variations across these to be probed directly from the underlying electronic-structure simulations.

\subsection{Barium Titanate}\label{batio3_subsect}

Finally, we turn to barium titanate (BaTiO$_3$), a ferroelectric perovskite used in capacitors and found in a rhombohedral phase at low temperatures, with increasing temperatures causing it to undergo transitions to orthorhombic, tetragonal, and cubic phases. Here, we will simulate its three most stable phases as different distortions to a regular cubic cell at 0 K~\bibnote{Structures have once again been loaded from the Materials Project database under the identifiers mp-5020, mp-5777, and mp-5986.}.\\

\begin{figure}[htbp]
    \centering
    \includegraphics[width=\textwidth]{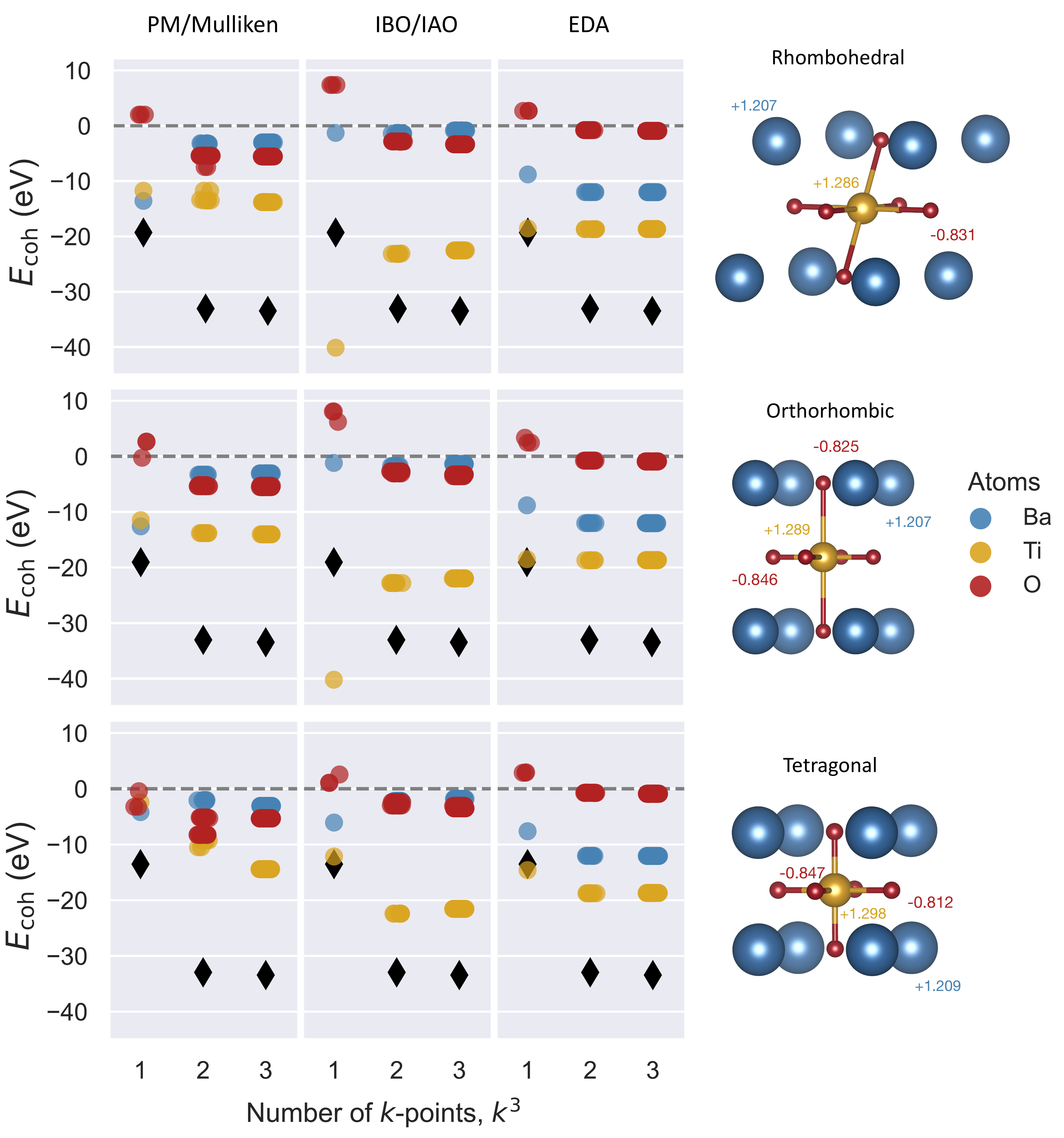}
    \caption{Atomic cohesive energies for different BaTiO$_3$ polymorphs. Black markers represent total cohesive energies, and the panel on the right shows IAO-based Mulliken charges.}
    \label{fig_4}
\end{figure}
The crystalline polymorphs of BaTiO$_3$ were recently studied in Ref. \citenum{dronskowski_jpcc_2021}. In that work, individual electrostatic potentials computed from atomic charges were found to be too constant across the phases for definitive conclusions to be drawn from differences among them. The sensitivity of the Madelung energies derived from said charges, on the other hand, was observed to improve significantly, but for these, regularity was observed to rule electrostatics with higher symmetries resulting in greater stabilizations. As such, while electrostatics alone predicted a preferred sequence of the polymorphs in clear favor of the most symmetric cubic phase, only total energies at zero temperature could confirm the experimental preference of the rhombohedral phase over the orthorhombic, tetragonal, and cubic counterparts.\\

Total cohesive energies of the three BaTiO$_3$ polymorphs exhibit only miniscule differences, owing to the similarity of the phases, and this picture is mirrored in the results of all MO- and AO-based partitioning schemes used herein, cf. Fig. \ref{fig_4}. Differences once again exist among the atomic cohesive energies derived from these, albeit significantly less so than what was observed for GaN in Fig. \ref{fig_3}. Whereas the extent of polarization between the phases of GaN was found to differ some, variations within the results of a given partitioning scheme are far fewer in the case of BaTiO$_3$ upon a sampling in $\bm{k}$-space. The net gain in stabilization is assigned to the Ti cation in all partitioning schemes, particularly so in the IBO/IAO results, while both MO-based schemes predict the energies associated with the Ba and O atoms to be practically the same in moving from vacuum to the crystal form. The AO-based EDA scheme, in contrast, predicts barium to be stabilized in all the three phases of BaTiO$_3$.\\

Using a so-called crystal orbital bond index, the authors of Ref. \citenum{dronskowski_jpcc_2021} identified the Ba--O bonds as ionic, while those between Ti and O were found to have some covalent bonding character. However, this picture is only vaguely repeated in the IAO-based partial charges of Fig. \ref{fig_4}, with significant deviations from the expected Ti$^{+4}$, Ba$^{+2}$, and O$^{-2}$ oxidation states~\bibnote{We note how our charges differ from those in Ref. \citenum{dronskowski_jpcc_2021}, but not by more than what is to be expected when considering the conceptual differences between Mulliken and L{\"o}wdin procedures for calculating these.}, nor do such trends emerge unambiguously from the atomic cohesive energies in Fig. \ref{fig_4}. While we will here refrain from discussing the general interplay between ionicity and covalency on the basis of atomic decompositions like the ones of the present work, the strong stabilization of the Ti cation and a corresponding lack of response to cohesion of both the Ba and O ions may instead point to somewhat more regular bonding patterns in this solid. However, more work beyond these early studies is needed in order to substantiate such claims any further.

\section{Summary and Conclusions}\label{summary_concl_sect}

By means of two atomic partitioning schemes---operating either on the basis of the spatial locality of crystalline AO basis functions or that of linear combinations of these within periodic systems---we have reported pilot results for decomposed cohesive energies at the KS-DFT level applied to both molecular polymers and three-dimensional solids. Beside exploring the robustness of both schemes under the introduction of periodic boundaries and effective core potentials, we have also examined what additional information may be drawn from probing local rather than global electronic structures in periodic matter. How well do these atomic contributions to cohesive properties align with prior expectations, and how may one seek to correlate changes in these with relative stabilizations of different crystalline polymorphs?\\

In general, we found that periodicity results in only minor changes to total and atomic cohesive energies, whereas the introduction of pseudopotentials can lead to more pronounced shifts. In line with recent simulations of chemical (non-periodic) Hamiltonians~\cite{eriksen_atom_energy_jctc_2023}, an MO-based partitioning scheme based on a combination of spatially localized IBOs and IAO-based Mulliken populations was found to yield the most consistent, uniform, and robust results, convincingly predicting net stabilizations within a periodic polymer of more electronegative constituent second-row atoms. For simulations of different GaN polymorphs, the scheme predicted a net stabilization of the more electronegative ion, again unlike AO-based alternatives, for which a consistent stabilization of all constituent atoms was predicted. Moreover, it was demonstrated that the anticipated error in atomic cohesive energies due to the use of pseudopotentials is effectively mitigated when comparing relative rather than absolute quantitites, e.g., energy changes across various phases. For investigations of surface reaction and adsorption profiles at an atomic level, these artefacts are thus expected to cancel out.\\ 

On the basis of the present pilot results, we believe comparisons of atomic cohesive energies simulated using our periodic implementations show promise as a new tool to be applied in elucidating relative phase stabilities in computational solid-state physics at zero temperature. The early applications studied herein aside, future work is required to provide sufficient grounds for interpretation of such atomic contributions to periodic properties in solids and establish the emergence of more general trends to be used in future applications. Beyond what is true for the simulation of molecules in isolation, the ability to probe local rather than global electronic structures in periodic matter further has the potential to shed light on truly fine-grained phenomena present in cells with certain vacancies, more disordered materials, or even the driving forces behind chemical reactions on solid surfaces. For such purposes, an optimized implementation of the MO-based decomposition scheme proposed here is under development in the {\texttt{CP2K}} software package~\cite{CP2K}, which will allow for more intriguing challenges at the interface of quantum chemistry and computational solid-state physics to be studied. Moving further beyond the present work, potential generalizations that would allow us to decompose total electronic energies directly in $\bm{k}$-space are of interest, for instance, through the construction of sets of crystal or Bloch-like IAOs and corresponding  Wannier functions~\cite{lehtola_jonsson_pmwf_jctc_2017,cui_period_dmet_jctc_2020,valeev_pmwf_jctc_2021,schaefer_iao_wannier_jcp_2021,zhu_tew_bloch_iao_arxiv_2024}.

\section*{Acknowledgments}

The authors are thankful to J{\"u}rg Hutter (University of Z{\"u}rich) for providing comments on an earlier draft of the present work and to Susi Lehtola (University of Helsinki) for providing comments on an initial preprint. This work was supported by two research grants awarded to JJE, no. 37411 from VILLUM FONDEN (a part of THE VELUX FOUNDATIONS) and no. 10.46540/2064-00007B from the Independent Research Fund Denmark.

\section*{Supporting Information}

The supporting information (SI) provides further details on the two AO- and MO-based decomposition schemes studied herein. Furthermore, Fig. S1 of the SI provides a convergence test of the cell volume in connection to the results in Fig. \ref{fig_1}, while Table S1 reports the individual atomic cohesive energies of Fig. \ref{fig_1}. The cumulative contributions to Fig. \ref{fig_1} as well as the effect of employing different minimal basis sets in the PBC/PP-based IBO/IAO decompositions are provided in Figs. S2 and S3 and Fig. S4, respectively. Table S2 reports the partial charges of the atoms in Fig. \ref{fig_2}, while the atomic cohesive energies of the PAN polymer simulated using either all-electron basis sets or pseudopotentials are compared in Fig. S5. Finally, PBC-based atomic cohesive energies of the four most stable phases of GaN computed either with or without PPs are compared to one another in Fig. S6.

\section*{Data Availability}

Data in support of the findings of this study are available within the article and its SI.


\providecommand{\latin}[1]{#1}
\makeatletter
\providecommand{\doi}
  {\begingroup\let\do\@makeother\dospecials
  \catcode`\{=1 \catcode`\}=2 \doi@aux}
\providecommand{\doi@aux}[1]{\endgroup\texttt{#1}}
\makeatother
\providecommand*\mcitethebibliography{\thebibliography}
\csname @ifundefined\endcsname{endmcitethebibliography}
  {\let\endmcitethebibliography\endthebibliography}{}


\begin{mcitethebibliography}{69}
\providecommand*\natexlab[1]{#1}
\providecommand*\mciteSetBstSublistMode[1]{}
\providecommand*\mciteSetBstMaxWidthForm[2]{}
\providecommand*\mciteBstWouldAddEndPuncttrue
  {\def\EndOfBibitem{\unskip.}}
\providecommand*\mciteBstWouldAddEndPunctfalse
  {\let\EndOfBibitem\relax}
\providecommand*\mciteSetBstMidEndSepPunct[3]{}
\providecommand*\mciteSetBstSublistLabelBeginEnd[3]{}
\providecommand*\EndOfBibitem{}
\mciteSetBstSublistMode{f}
\mciteSetBstMaxWidthForm{subitem}{(\alph{mcitesubitemcount})}
\mciteSetBstSublistLabelBeginEnd
  {\mcitemaxwidthsubitemform\space}
  {\relax}
  {\relax}

\bibitem[Monkhorst and Pack(1976)Monkhorst, and Pack]{Monkhorst-Pack_kmesh}
Monkhorst,~H.~J.; Pack,~J.~D. Special Points for Brillouin-Zone Integrations.
  \emph{Phys. Rev. B} \textbf{1976}, \emph{13}, 5188\relax
\mciteBstWouldAddEndPuncttrue
\mciteSetBstMidEndSepPunct{\mcitedefaultmidpunct}
{\mcitedefaultendpunct}{\mcitedefaultseppunct}\relax
\EndOfBibitem
\bibitem[Eriksen(2020)]{eriksen_decodense_jcp_2020}
Eriksen,~J.~J. {Mean-Field Density Matrix Decompositions}. \emph{{J}. {C}hem.
  {P}hys.} \textbf{2020}, \emph{153}, 214109\relax
\mciteBstWouldAddEndPuncttrue
\mciteSetBstMidEndSepPunct{\mcitedefaultmidpunct}
{\mcitedefaultendpunct}{\mcitedefaultseppunct}\relax
\EndOfBibitem
\bibitem[Eriksen(2022)]{eriksen_elec_ex_decomp_jcp_2022}
Eriksen,~J.~J. {Electronic Excitations Through the Prism of Mean-Field
  Decomposition Techniques}. \emph{{J}. {C}hem. {P}hys.} \textbf{2022},
  \emph{156}, 061101\relax
\mciteBstWouldAddEndPuncttrue
\mciteSetBstMidEndSepPunct{\mcitedefaultmidpunct}
{\mcitedefaultendpunct}{\mcitedefaultseppunct}\relax
\EndOfBibitem
\bibitem[Eriksen(2021)]{eriksen_local_condensed_phase_jpcl_2021}
Eriksen,~J.~J. {Decomposed Mean-Field Simulations of Local Properties in
  Condensed Phases}. \emph{J. Phys. Chem. Lett.} \textbf{2021}, \emph{12},
  6048\relax
\mciteBstWouldAddEndPuncttrue
\mciteSetBstMidEndSepPunct{\mcitedefaultmidpunct}
{\mcitedefaultendpunct}{\mcitedefaultseppunct}\relax
\EndOfBibitem
\bibitem[Kjeldal and Eriksen(2023)Kjeldal, and
  Eriksen]{eriksen_nn_qm7_decomp_jctc_2023}
Kjeldal,~F.~{\O}.; Eriksen,~J.~J. {Decomposing Chemical Space: Applications to
  the Machine Learning of Atomic Energies}. \emph{{J}. {C}hem. {T}heory
  {C}omput.} \textbf{2023}, \emph{19}, 2029\relax
\mciteBstWouldAddEndPuncttrue
\mciteSetBstMidEndSepPunct{\mcitedefaultmidpunct}
{\mcitedefaultendpunct}{\mcitedefaultseppunct}\relax
\EndOfBibitem
\bibitem[Kjeldal and Eriksen(2023)Kjeldal, and
  Eriksen]{eriksen_atom_energy_jctc_2023}
Kjeldal,~F.~{\O}.; Eriksen,~J.~J. {Properties of Local Electronic Structures}.
  \emph{{J}. {C}hem. {T}heory {C}omput.} \textbf{2023}, \emph{19}, 9228\relax
\mciteBstWouldAddEndPuncttrue
\mciteSetBstMidEndSepPunct{\mcitedefaultmidpunct}
{\mcitedefaultendpunct}{\mcitedefaultseppunct}\relax
\EndOfBibitem
\bibitem[Helgaker \latin{et~al.}(2000)Helgaker, J{\o}rgensen, and Olsen]{mest}
Helgaker,~T.; J{\o}rgensen,~P.; Olsen,~J. \emph{{M}olecular
  {E}lectronic-{S}tructure {T}heory}; Wiley \& Sons, Ltd.: West Sussex, UK,
  2000\relax
\mciteBstWouldAddEndPuncttrue
\mciteSetBstMidEndSepPunct{\mcitedefaultmidpunct}
{\mcitedefaultendpunct}{\mcitedefaultseppunct}\relax
\EndOfBibitem
\bibitem[Martin(2004)]{martin_elec_struct}
Martin,~R.~M. \emph{Electronic Structure: Basic Theory and Practical Methods};
  Cambridge University Press: Cambridge, UK, 2004\relax
\mciteBstWouldAddEndPuncttrue
\mciteSetBstMidEndSepPunct{\mcitedefaultmidpunct}
{\mcitedefaultendpunct}{\mcitedefaultseppunct}\relax
\EndOfBibitem
\bibitem[Chulhai and Goodpaster(2018)Chulhai, and
  Goodpaster]{goodpaster_embedding_jctc_2018}
Chulhai,~D.~V.; Goodpaster,~J.~D. Projection-Based Correlated Wave Function in
  Density Functional Theory Embedding for Periodic Systems. \emph{J. Chem.
  Theory Comput.} \textbf{2018}, \emph{14}, 1928\relax
\mciteBstWouldAddEndPuncttrue
\mciteSetBstMidEndSepPunct{\mcitedefaultmidpunct}
{\mcitedefaultendpunct}{\mcitedefaultseppunct}\relax
\EndOfBibitem
\bibitem[Sanchez-Portal \latin{et~al.}(1995)Sanchez-Portal, Artacho, and
  Soler]{soler_projection_sscomm_1995}
Sanchez-Portal,~D.; Artacho,~E.; Soler,~J.~M. Projection of Plane-Wave
  Calculations into Atomic Orbitals. \emph{Solid State Commun.} \textbf{1995},
  \emph{95}, 685\relax
\mciteBstWouldAddEndPuncttrue
\mciteSetBstMidEndSepPunct{\mcitedefaultmidpunct}
{\mcitedefaultendpunct}{\mcitedefaultseppunct}\relax
\EndOfBibitem
\bibitem[Maintz \latin{et~al.}(2013)Maintz, Deringer, Tchougreeff, and
  Dronskowski]{dronskowski_jcc_2013}
Maintz,~S.; Deringer,~V.~L.; Tchougreeff,~A.~L.; Dronskowski,~R. Analytic
  Projection from Plane-Wave and PAW Wavefunctions and Application to
  Chemical-Bonding Analysis in Solids. \emph{J. Comput. Chem.} \textbf{2013},
  \emph{34}, 2557\relax
\mciteBstWouldAddEndPuncttrue
\mciteSetBstMidEndSepPunct{\mcitedefaultmidpunct}
{\mcitedefaultendpunct}{\mcitedefaultseppunct}\relax
\EndOfBibitem
\bibitem[Maintz \latin{et~al.}(2016)Maintz, Deringer, Tchougreeff, and
  Dronskowski]{lobster_jcc_2016}
Maintz,~S.; Deringer,~V.~L.; Tchougreeff,~A.~L.; Dronskowski,~R.
  {\texttt{LOBSTER}}: A Tool to Extract Chemical Bonding from Plane-Wave Based
  DFT. \emph{J. Comput. Chem.} \textbf{2016}, \emph{37}, 1030\relax
\mciteBstWouldAddEndPuncttrue
\mciteSetBstMidEndSepPunct{\mcitedefaultmidpunct}
{\mcitedefaultendpunct}{\mcitedefaultseppunct}\relax
\EndOfBibitem
\bibitem[Nelson \latin{et~al.}(2020)Nelson, Ertural, George, Deringer, Hautier,
  and Dronskowski]{lobster_jcc_2020}
Nelson,~R.; Ertural,~C.; George,~J.; Deringer,~V.~L.; Hautier,~G.;
  Dronskowski,~R. {\texttt{LOBSTER}}: Local Orbital Projections, Atomic
  Charges, and Chemical-Bonding Analysis from Projector-Augmented Wave-Based
  Density-Functional Theory. \emph{J. Comput. Chem.} \textbf{2020}, \emph{41},
  1931\relax
\mciteBstWouldAddEndPuncttrue
\mciteSetBstMidEndSepPunct{\mcitedefaultmidpunct}
{\mcitedefaultendpunct}{\mcitedefaultseppunct}\relax
\EndOfBibitem
\bibitem[Marzari \latin{et~al.}(2012)Marzari, Mostofi, Yates, Souza, and
  Vanderbilt]{vanderbilt_wannier_rmp_2012}
Marzari,~N.; Mostofi,~A.~A.; Yates,~J.~R.; Souza,~I.; Vanderbilt,~D. Maximally
  Localized Wannier Functions: Theory and Applications. \emph{Rev. Mod. Phys.}
  \textbf{2012}, \emph{84}, 1419\relax
\mciteBstWouldAddEndPuncttrue
\mciteSetBstMidEndSepPunct{\mcitedefaultmidpunct}
{\mcitedefaultendpunct}{\mcitedefaultseppunct}\relax
\EndOfBibitem
\bibitem[Pauls \latin{et~al.}(2023)Pauls, Schnieders, and
  Dronskowski]{dronskowski_lmo_jpca_2023}
Pauls,~M.; Schnieders,~D.; Dronskowski,~R. Embedded Localized Molecular-Orbital
  Representations for Periodic Wave Functions. \emph{J. Phys. Chem. A}
  \textbf{2023}, \emph{127}, 6541\relax
\mciteBstWouldAddEndPuncttrue
\mciteSetBstMidEndSepPunct{\mcitedefaultmidpunct}
{\mcitedefaultendpunct}{\mcitedefaultseppunct}\relax
\EndOfBibitem
\bibitem[Kato \latin{et~al.}(2017)Kato, Hongo, Maezono, Higashi, Kunioku,
  Yabuuchi, Suzuki, Okajima, Zhong, Nakano, Abe, and
  Kageyama]{kageyama_jacs_2017}
Kato,~D.; Hongo,~K.; Maezono,~R.; Higashi,~M.; Kunioku,~H.; Yabuuchi,~M.;
  Suzuki,~H.; Okajima,~H.; Zhong,~C.; Nakano,~K.; Abe,~R.; Kageyama,~H. Valence
  Band Engineering of Layered Bismuth Oxyhalides toward Stable Visible-Light
  Water Splitting: Madelung Site Potential Analysis. \emph{J. Am. Chem. Soc.}
  \textbf{2017}, \emph{139}, 18725\relax
\mciteBstWouldAddEndPuncttrue
\mciteSetBstMidEndSepPunct{\mcitedefaultmidpunct}
{\mcitedefaultendpunct}{\mcitedefaultseppunct}\relax
\EndOfBibitem
\bibitem[Luo \latin{et~al.}(2021)Luo, Qiao, and
  Dronskowski]{dronskowski_acie_2021}
Luo,~D.; Qiao,~X.; Dronskowski,~R. Predicting Nitrogen-Based Families of
  Compounds: Transition-Metal Guanidinates TCN$_3$ (T = V, Nb, Ta) and
  Ortho-Nitrido Carbonates T$^{\prime}_2$CN$_4$ (T$^{\prime}$ = Ti, Zr, Hf).
  \emph{Angew. Chem. Int. Ed.} \textbf{2021}, \emph{60}, 486\relax
\mciteBstWouldAddEndPuncttrue
\mciteSetBstMidEndSepPunct{\mcitedefaultmidpunct}
{\mcitedefaultendpunct}{\mcitedefaultseppunct}\relax
\EndOfBibitem
\bibitem[Hughbanks and Hoffmann(1983)Hughbanks, and
  Hoffmann]{hughbanks_hoffmann_jacs_1983}
Hughbanks,~T.; Hoffmann,~R. Chains of Trans-Edge-Sharing Molybdenum Octahedra:
  Metal-Metal Bonding in Extended Systems. \emph{J. Am. Chem. Soc.}
  \textbf{1983}, \emph{105}, 3528\relax
\mciteBstWouldAddEndPuncttrue
\mciteSetBstMidEndSepPunct{\mcitedefaultmidpunct}
{\mcitedefaultendpunct}{\mcitedefaultseppunct}\relax
\EndOfBibitem
\bibitem[Dronskowski and Bl{\"o}chl(1993)Dronskowski, and
  Bl{\"o}chl]{dronskowski_blochl_jpc_1993}
Dronskowski,~R.; Bl{\"o}chl,~P.~E. Crystal Orbital Hamilton Populations (COHP):
  Energy-Resolved Visualization of Chemical Bonding in Solids Based on
  Density-Functional Calculations. \emph{J. Phys. Chem.} \textbf{1993},
  \emph{97}, 8617\relax
\mciteBstWouldAddEndPuncttrue
\mciteSetBstMidEndSepPunct{\mcitedefaultmidpunct}
{\mcitedefaultendpunct}{\mcitedefaultseppunct}\relax
\EndOfBibitem
\bibitem[Deringer \latin{et~al.}(2011)Deringer, Tchougreeff, and
  Dronskowski]{dronskowski_jpca_2011}
Deringer,~V.~L.; Tchougreeff,~A.~L.; Dronskowski,~R. Crystal Orbital Hamilton
  Population (COHP) Analysis as Projected from Plane-Wave Basis Sets. \emph{J.
  Phys. Chem. A} \textbf{2011}, \emph{115}, 5461\relax
\mciteBstWouldAddEndPuncttrue
\mciteSetBstMidEndSepPunct{\mcitedefaultmidpunct}
{\mcitedefaultendpunct}{\mcitedefaultseppunct}\relax
\EndOfBibitem
\bibitem[M{\"u}ller \latin{et~al.}(2021)M{\"u}ller, Ertural, Hempelmann, and
  Dronskowski]{dronskowski_jpcc_2021}
M{\"u}ller,~P.~C.; Ertural,~C.; Hempelmann,~J.; Dronskowski,~R. Crystal Orbital
  Bond Index: Covalent Bond Orders in Solids. \emph{J. Phys. Chem. C}
  \textbf{2021}, \emph{125}, 7959\relax
\mciteBstWouldAddEndPuncttrue
\mciteSetBstMidEndSepPunct{\mcitedefaultmidpunct}
{\mcitedefaultendpunct}{\mcitedefaultseppunct}\relax
\EndOfBibitem
\bibitem[George \latin{et~al.}(2022)George, Petretto, Naik, Esters, Jackson,
  Nelson, Dronskowski, Rignanese, and Hautier]{george_hautier_cpc_2022}
George,~J.; Petretto,~G.; Naik,~A.; Esters,~M.; Jackson,~A.~J.; Nelson,~R.;
  Dronskowski,~R.; Rignanese,~M.; Hautier,~G. Automated Bonding Analysis with
  Crystal Orbital Hamilton Populations. \emph{ChemPlusChem} \textbf{2022},
  \emph{87}, e202200123\relax
\mciteBstWouldAddEndPuncttrue
\mciteSetBstMidEndSepPunct{\mcitedefaultmidpunct}
{\mcitedefaultendpunct}{\mcitedefaultseppunct}\relax
\EndOfBibitem
\bibitem[Ertural \latin{et~al.}(2019)Ertural, Steinberg, and
  Dronskowski]{dronskowski_rsc_adv_2019}
Ertural,~C.; Steinberg,~S.; Dronskowski,~R. Development of a Robust Tool to
  Extract Mulliken and L{\"o}wdin Charges from Plane Waves and its Application
  to Solid-State Materials. \emph{RSC Adv.} \textbf{2019}, \emph{9},
  29821\relax
\mciteBstWouldAddEndPuncttrue
\mciteSetBstMidEndSepPunct{\mcitedefaultmidpunct}
{\mcitedefaultendpunct}{\mcitedefaultseppunct}\relax
\EndOfBibitem
\bibitem[Knizia(2013)]{knizia_iao_ibo_jctc_2013}
Knizia,~G. {Intrinsic Atomic Orbitals: An Unbiased Bridge Between Quantum
  Theory and Chemical Concepts}. \emph{{J}. {C}hem. {T}heory {C}omput.}
  \textbf{2013}, \emph{9}, 4834\relax
\mciteBstWouldAddEndPuncttrue
\mciteSetBstMidEndSepPunct{\mcitedefaultmidpunct}
{\mcitedefaultendpunct}{\mcitedefaultseppunct}\relax
\EndOfBibitem
\bibitem[Janowski(2014)]{janowski_quambo_jctc_2014}
Janowski,~T. {Near Equivalence of Intrinsic Atomic Orbitals and Quasiatomic
  Orbitals}. \emph{{J}. {C}hem. {T}heory {C}omput.} \textbf{2014}, \emph{10},
  3085\relax
\mciteBstWouldAddEndPuncttrue
\mciteSetBstMidEndSepPunct{\mcitedefaultmidpunct}
{\mcitedefaultendpunct}{\mcitedefaultseppunct}\relax
\EndOfBibitem
\bibitem[Lehtola and J{\'o}nsson(2014)Lehtola, and
  J{\'o}nsson]{lehtola_jonsson_pm_jctc_2014}
Lehtola,~S.; J{\'o}nsson,~H. {Pipek-Mezey Orbital Localization Using Various
  Partial Charge Estimates}. \emph{{J}. {C}hem. {T}heory {C}omput.}
  \textbf{2014}, \emph{10}, 642\relax
\mciteBstWouldAddEndPuncttrue
\mciteSetBstMidEndSepPunct{\mcitedefaultmidpunct}
{\mcitedefaultendpunct}{\mcitedefaultseppunct}\relax
\EndOfBibitem
\bibitem[Nakai(2002)]{nakai_eda_partitioning_cpl_2002}
Nakai,~H. {Energy Density Analysis with Kohn-Sham Orbitals}. \emph{Chem. Phys.
  Lett.} \textbf{2002}, \emph{363}, 73\relax
\mciteBstWouldAddEndPuncttrue
\mciteSetBstMidEndSepPunct{\mcitedefaultmidpunct}
{\mcitedefaultendpunct}{\mcitedefaultseppunct}\relax
\EndOfBibitem
\bibitem[Kikuchi \latin{et~al.}(2009)Kikuchi, Imamura, and
  Nakai]{nakai_eda_partitioning_ijqc_2009}
Kikuchi,~Y.; Imamura,~Y.; Nakai,~H. {One-Body Energy Decomposition Schemes
  Revisited: Assessment of Mulliken-, Grid-, and Conventional Energy Density
  Analyses}. \emph{Int. J. Quantum Chem.} \textbf{2009}, \emph{109}, 2464\relax
\mciteBstWouldAddEndPuncttrue
\mciteSetBstMidEndSepPunct{\mcitedefaultmidpunct}
{\mcitedefaultendpunct}{\mcitedefaultseppunct}\relax
\EndOfBibitem
\bibitem[Baba \latin{et~al.}(2006)Baba, Takeuchi, and Nakai]{baba2006natural}
Baba,~T.; Takeuchi,~M.; Nakai,~H. Natural Atomic Orbital Based Energy Density
  Analysis: Implementation and Applications. \emph{Chem. Phys. Lett.}
  \textbf{2006}, \emph{424}, 193\relax
\mciteBstWouldAddEndPuncttrue
\mciteSetBstMidEndSepPunct{\mcitedefaultmidpunct}
{\mcitedefaultendpunct}{\mcitedefaultseppunct}\relax
\EndOfBibitem
\bibitem[Imamura \latin{et~al.}(2007)Imamura, Takahashi, and
  Nakai]{imamura2007grid}
Imamura,~Y.; Takahashi,~A.; Nakai,~H. Grid-Based Energy Density Analysis:
  Implementation and Assessment. \emph{J. Chem. Phys.} \textbf{2007},
  \emph{126}, 034103\relax
\mciteBstWouldAddEndPuncttrue
\mciteSetBstMidEndSepPunct{\mcitedefaultmidpunct}
{\mcitedefaultendpunct}{\mcitedefaultseppunct}\relax
\EndOfBibitem
\bibitem[Goedecker \latin{et~al.}(1996)Goedecker, Teter, and Hutter]{GTH_pp}
Goedecker,~S.; Teter,~M.; Hutter,~J. Separable Dual-Space Gaussian
  Pseudopotentials. \emph{Phys. Rev. B: Condens. Matter Mater. Phys.}
  \textbf{1996}, \emph{54}, 1703\relax
\mciteBstWouldAddEndPuncttrue
\mciteSetBstMidEndSepPunct{\mcitedefaultmidpunct}
{\mcitedefaultendpunct}{\mcitedefaultseppunct}\relax
\EndOfBibitem
\bibitem[Hartwigsen \latin{et~al.}(1998)Hartwigsen, Goedecker, and
  Hutter]{GTH_pp_rel}
Hartwigsen,~C.; Goedecker,~S.; Hutter,~J. Relativistic Separable Dual-Space
  Gaussian Pseudopotentials from H to Rn. \emph{Phys. Rev. B: Condens. Matter
  Mater. Phys.} \textbf{1998}, \emph{58}, 3641\relax
\mciteBstWouldAddEndPuncttrue
\mciteSetBstMidEndSepPunct{\mcitedefaultmidpunct}
{\mcitedefaultendpunct}{\mcitedefaultseppunct}\relax
\EndOfBibitem
\bibitem[McClain \latin{et~al.}(2017)McClain, Sun, Chan, and
  Berkelbach]{berkelbach_eom_cc_jctc_2017}
McClain,~J.; Sun,~Q.; Chan,~G. K.-L.; Berkelbach,~T.~C. Gaussian-Based
  Coupled-Cluster Theory for the Ground-State and Band Structure of Solids.
  \emph{J. Chem. Theory Comput.} \textbf{2017}, \emph{13}, 1209\relax
\mciteBstWouldAddEndPuncttrue
\mciteSetBstMidEndSepPunct{\mcitedefaultmidpunct}
{\mcitedefaultendpunct}{\mcitedefaultseppunct}\relax
\EndOfBibitem
\bibitem[Sun \latin{et~al.}(2018)Sun, Berkelbach, Blunt, Booth, Guo, Li, Liu,
  McClain, Sayfutyarova, Sharma, Wouters, and Chan]{pyscf_wires_2018}
Sun,~Q.; Berkelbach,~T.~C.; Blunt,~N.~S.; Booth,~G.~H.; Guo,~S.; Li,~Z.;
  Liu,~J.; McClain,~J.~D.; Sayfutyarova,~E.~R.; Sharma,~S.; Wouters,~S.;
  Chan,~G. K.-L. {{\texttt{PySCF}}: The Python-Based Simulations of Chemistry
  Framework}. \emph{{W}IREs {C}omput. {M}ol. {S}ci.} \textbf{2018}, \emph{8},
  e1340\relax
\mciteBstWouldAddEndPuncttrue
\mciteSetBstMidEndSepPunct{\mcitedefaultmidpunct}
{\mcitedefaultendpunct}{\mcitedefaultseppunct}\relax
\EndOfBibitem
\bibitem[Sun \latin{et~al.}(2020)Sun, Zhang, Banerjee, Bao, Barbry, Blunt,
  Bogdanov, Booth, Chen, Cui, Eriksen, Gao, Guo, Hermann, Hermes, Koh, Koval,
  Lehtola, Li, Liu, Mardirossian, McClain, Motta, Mussard, Pham, Pulkin,
  Purwanto, Robinson, Ronca, Sayfutyarova, Scheurer, Schurkus, Smith, Sun, Sun,
  Upadhyay, Wagner, Wang, White, Whitfield, Williamson, Wouters, Yang, Yu, Zhu,
  Berkelbach, Sharma, Sokolov, and Chan]{pyscf_jcp_2020}
Sun,~Q.; Zhang,~X.; Banerjee,~S.; Bao,~P.; Barbry,~M.; Blunt,~N.~S.;
  Bogdanov,~N.~A.; Booth,~G.~H.; Chen,~J.; Cui,~Z.-H.; Eriksen,~J.~J.; Gao,~Y.;
  Guo,~S.; Hermann,~J.; Hermes,~M.~R.; Koh,~K.; Koval,~P.; Lehtola,~S.; Li,~Z.;
  Liu,~J.; Mardirossian,~N.; McClain,~J.~D.; Motta,~M.; Mussard,~B.;
  Pham,~H.~Q.; Pulkin,~A.; Purwanto,~W.; Robinson,~P.~J.; Ronca,~E.;
  Sayfutyarova,~E.~R.; Scheurer,~M.; Schurkus,~H.~F.; Smith,~J. E.~T.; Sun,~C.;
  Sun,~S.-N.; Upadhyay,~S.; Wagner,~L.~K.; Wang,~X.; White,~A.;
  Whitfield,~J.~D.; Williamson,~M.~J.; Wouters,~S.; Yang,~J.; Yu,~J.~M.;
  Zhu,~T.; Berkelbach,~T.~C.; Sharma,~S.; Sokolov,~A.~Y.; Chan,~G. K.-L.
  {Recent Developments in the {\texttt{PySCF}} Program Package}. \emph{J. Chem.
  Phys.} \textbf{2020}, \emph{153}, 024109\relax
\mciteBstWouldAddEndPuncttrue
\mciteSetBstMidEndSepPunct{\mcitedefaultmidpunct}
{\mcitedefaultendpunct}{\mcitedefaultseppunct}\relax
\EndOfBibitem
\bibitem[Sun \latin{et~al.}(2017)Sun, Berkelbach, McClain, and
  Chan]{GDF_MDF_PySCF}
Sun,~Q.; Berkelbach,~T.~C.; McClain,~J.~D.; Chan,~G. K.-L. Gaussian and
  Plane-Wave Mixed Density Fitting for Periodic Systems. \emph{J. Chem. Phys.}
  \textbf{2017}, \emph{147}, 164119\relax
\mciteBstWouldAddEndPuncttrue
\mciteSetBstMidEndSepPunct{\mcitedefaultmidpunct}
{\mcitedefaultendpunct}{\mcitedefaultseppunct}\relax
\EndOfBibitem
\bibitem[VandeVondele \latin{et~al.}(2005)VandeVondele, Krack, Mohamed,
  Parrinello, Chassaing, and Hutter]{CP2K_GPW_also_pyscf}
VandeVondele,~J.; Krack,~M.; Mohamed,~F.; Parrinello,~M.; Chassaing,~T.;
  Hutter,~J. Quickstep: Fast and Accurate Density Functional Calculations Using
  a Mixed Gaussian and Plane Waves Approach. \emph{Comp. Phys. Comm.}
  \textbf{2005}, \emph{167}, 103\relax
\mciteBstWouldAddEndPuncttrue
\mciteSetBstMidEndSepPunct{\mcitedefaultmidpunct}
{\mcitedefaultendpunct}{\mcitedefaultseppunct}\relax
\EndOfBibitem
\bibitem[K{\"u}hne \latin{et~al.}(2020)K{\"u}hne, Iannuzzi, Del~Ben, Rybkin,
  Seewald, Stein, Laino, Khaliullin, Sch{\"u}tt, Schiffmann, Golze, Wilhelm,
  Chulkov, Bani-Hashemian, Weber, Bor{\v{s}}tnik, Taillefumier, Jakobovits,
  Lazzaro, Pabst, M{\"u}ller, Schade, Guidon, Andermatt, Holmberg, Schenter,
  Hehn, Bussy, Belleflamme, Tabacchi, Gl{\"o}{\ss}, Lass, Bethune, Mundy,
  Plessl, Watkins, VandeVondele, Krack, and Hutter]{CP2K}
K{\"u}hne,~T.~D.; Iannuzzi,~M.; Del~Ben,~M.; Rybkin,~V.~V.; Seewald,~P.;
  Stein,~F.; Laino,~T.; Khaliullin,~R.~Z.; Sch{\"u}tt,~O.; Schiffmann,~F.;
  Golze,~D.; Wilhelm,~J.; Chulkov,~S.; Bani-Hashemian,~M.~H.; Weber,~V.;
  Bor{\v{s}}tnik,~U.; Taillefumier,~M.; Jakobovits,~A.~S.; Lazzaro,~A.;
  Pabst,~H.; M{\"u}ller,~T.; Schade,~R.; Guidon,~M.; Andermatt,~S.;
  Holmberg,~N.; Schenter,~G.~K.; Hehn,~A.; Bussy,~A.; Belleflamme,~F.;
  Tabacchi,~G.; Gl{\"o}{\ss},~A.; Lass,~M.; Bethune,~I.; Mundy,~C.~J.;
  Plessl,~C.; Watkins,~M.; VandeVondele,~J.; Krack,~M.; Hutter,~J.
  {\texttt{CP2K}}: An Electronic Structure and Molecular Dynamics Software
  Package -- Quickstep: Efficient and Accurate Electronic Structure
  Calculations. \emph{J. Chem. Phys.} \textbf{2020}, \emph{152}, 194103\relax
\mciteBstWouldAddEndPuncttrue
\mciteSetBstMidEndSepPunct{\mcitedefaultmidpunct}
{\mcitedefaultendpunct}{\mcitedefaultseppunct}\relax
\EndOfBibitem
\bibitem[Ewald(1921)]{ewald_sum}
Ewald,~P.~P. Die Berechnung Optischer und Elektrostatischer Gitterpotentiale.
  \emph{Ann. Phys. (Berl.)} \textbf{1921}, \emph{369}, 253\relax
\mciteBstWouldAddEndPuncttrue
\mciteSetBstMidEndSepPunct{\mcitedefaultmidpunct}
{\mcitedefaultendpunct}{\mcitedefaultseppunct}\relax
\EndOfBibitem
\bibitem[Perdew \latin{et~al.}(1996)Perdew, Burke, and
  Ernzerhof]{perdew_burke_ernzerhof_pbe_functional_prl_1996}
Perdew,~J.~P.; Burke,~K.; Ernzerhof,~M. {Generalized Gradient Approximation
  Made Simple}. \emph{Phys. Rev. Lett.} \textbf{1996}, \emph{77}, 3865\relax
\mciteBstWouldAddEndPuncttrue
\mciteSetBstMidEndSepPunct{\mcitedefaultmidpunct}
{\mcitedefaultendpunct}{\mcitedefaultseppunct}\relax
\EndOfBibitem
\bibitem[Becke(1988)]{B_lyp}
Becke,~A.~D. Density-Functional Exchange-Energy Approximation With Correct
  Asymptotic Behavior. \emph{Phys. Rev. A} \textbf{1988}, \emph{38}, 3098\relax
\mciteBstWouldAddEndPuncttrue
\mciteSetBstMidEndSepPunct{\mcitedefaultmidpunct}
{\mcitedefaultendpunct}{\mcitedefaultseppunct}\relax
\EndOfBibitem
\bibitem[Lee \latin{et~al.}(1988)Lee, Yang, and Parr]{b_L_yp}
Lee,~C.; Yang,~W.; Parr,~R.~G. Development of the Colle-Salvetti
  Correlation-Energy Formula into a Functional of the Electron Density.
  \emph{Phys. Rev. B} \textbf{1988}, \emph{37}, 785\relax
\mciteBstWouldAddEndPuncttrue
\mciteSetBstMidEndSepPunct{\mcitedefaultmidpunct}
{\mcitedefaultendpunct}{\mcitedefaultseppunct}\relax
\EndOfBibitem
\bibitem[Miehlich \latin{et~al.}(1989)Miehlich, Savin, Stoll, and
  Preuss]{bl_YP}
Miehlich,~B.; Savin,~A.; Stoll,~H.; Preuss,~H. Results Obtained With the
  Correlation Energy Density Functionals of Becke and Lee, Yang and Parr.
  \emph{Chem. Phys. Lett.} \textbf{1989}, \emph{157}, 200\relax
\mciteBstWouldAddEndPuncttrue
\mciteSetBstMidEndSepPunct{\mcitedefaultmidpunct}
{\mcitedefaultendpunct}{\mcitedefaultseppunct}\relax
\EndOfBibitem
\bibitem[Perdew \latin{et~al.}(2008)Perdew, Ruzsinszky, Csonka, Vydrov,
  Scuseria, Constantin, Zhou, and Burke]{PBEsol1}
Perdew,~J.~P.; Ruzsinszky,~A.; Csonka,~G.~I.; Vydrov,~O.~A.; Scuseria,~G.~E.;
  Constantin,~L.~A.; Zhou,~X.; Burke,~K. Restoring the Density-Gradient
  Expansion for Exchange in Solids and Surfaces. \emph{Phys. Rev. Lett.}
  \textbf{2008}, \emph{100}, 136406\relax
\mciteBstWouldAddEndPuncttrue
\mciteSetBstMidEndSepPunct{\mcitedefaultmidpunct}
{\mcitedefaultendpunct}{\mcitedefaultseppunct}\relax
\EndOfBibitem
\bibitem[Constantin \latin{et~al.}(2009)Constantin, Perdew, and
  Pitarke]{PBEsol2}
Constantin,~L.~A.; Perdew,~J.~P.; Pitarke,~J.~M. Exchange-Correlation Hole of a
  Generalized Gradient Approximation for Solids and Surfaces. \emph{Phys. Rev.
  B} \textbf{2009}, \emph{79}, 075126\relax
\mciteBstWouldAddEndPuncttrue
\mciteSetBstMidEndSepPunct{\mcitedefaultmidpunct}
{\mcitedefaultendpunct}{\mcitedefaultseppunct}\relax
\EndOfBibitem
\bibitem[Lehtola \latin{et~al.}(2018)Lehtola, Steigemann, Oliveira, and
  Marques]{libxc_software_x_2018}
Lehtola,~S.; Steigemann,~C.; Oliveira,~M. J.~T.; Marques,~M. A.~L. {Recent
  Developments in {\texttt{Libxc}} --- A Comprehensive Library of Functionals
  for Density Functional Theory}. \emph{Software X} \textbf{2018}, \emph{7},
  1\relax
\mciteBstWouldAddEndPuncttrue
\mciteSetBstMidEndSepPunct{\mcitedefaultmidpunct}
{\mcitedefaultendpunct}{\mcitedefaultseppunct}\relax
\EndOfBibitem
\bibitem[Godbout \latin{et~al.}(1992)Godbout, Salahub, Andzelm, and
  Wimmer]{DZVP}
Godbout,~N.; Salahub,~D.~R.; Andzelm,~J.; Wimmer,~E. Optimization of
  Gaussian-Type Basis Sets for Local Spin Density Functional Calculations. Part
  I. Boron Through Neon, Optimization Technique and Validation. \emph{Can. J.
  Chem.} \textbf{1992}, \emph{70}, 560\relax
\mciteBstWouldAddEndPuncttrue
\mciteSetBstMidEndSepPunct{\mcitedefaultmidpunct}
{\mcitedefaultendpunct}{\mcitedefaultseppunct}\relax
\EndOfBibitem
\bibitem[VandeVondele and Hutter(2007)VandeVondele, and
  Hutter]{gth_molopt_basis}
VandeVondele,~J.; Hutter,~J. Gaussian Basis Sets for Accurate Calculations on
  Molecular Systems in Gas and Condensed Phases. \emph{{J}. {C}hem. {P}hys.}
  \textbf{2007}, \emph{127}, 114105\relax
\mciteBstWouldAddEndPuncttrue
\mciteSetBstMidEndSepPunct{\mcitedefaultmidpunct}
{\mcitedefaultendpunct}{\mcitedefaultseppunct}\relax
\EndOfBibitem
\bibitem[Ditchfield \latin{et~al.}(1971)Ditchfield, Hehre, and
  Pople]{631Gs_H_ditchfield1971a}
Ditchfield,~R.; Hehre,~W.~J.; Pople,~J.~A. Self-Consistent Molecular-Orbital
  Methods. IX. An Extended Gaussian-Type Basis for Molecular-Orbital Studies of
  Organic Molecules. \emph{J. Chem. Phys.} \textbf{1971}, \emph{54},
  724--728\relax
\mciteBstWouldAddEndPuncttrue
\mciteSetBstMidEndSepPunct{\mcitedefaultmidpunct}
{\mcitedefaultendpunct}{\mcitedefaultseppunct}\relax
\EndOfBibitem
\bibitem[Hehre \latin{et~al.}(1972)Hehre, Ditchfield, and
  Pople]{631Gs_CNO_hehre1972a}
Hehre,~W.~J.; Ditchfield,~R.; Pople,~J.~A. Self-Consistent Molecular Orbital
  Methods. XII. Further Extensions of Gaussian-Type Basis Sets for Use in
  Molecular Orbital Studies of Organic Molecules. \emph{J. Chem. Phys.}
  \textbf{1972}, \emph{56}, 2257--2261\relax
\mciteBstWouldAddEndPuncttrue
\mciteSetBstMidEndSepPunct{\mcitedefaultmidpunct}
{\mcitedefaultendpunct}{\mcitedefaultseppunct}\relax
\EndOfBibitem
\bibitem[Hariharan and Pople(1973)Hariharan, and
  Pople]{631Gs_CNO_hariharan1973a}
Hariharan,~P.~C.; Pople,~J.~A. The Influence of Polarization Functions on
  Molecular Orbital Hydrogenation Energies. \emph{Theor. Chim. Acta}
  \textbf{1973}, \emph{28}, 213--222\relax
\mciteBstWouldAddEndPuncttrue
\mciteSetBstMidEndSepPunct{\mcitedefaultmidpunct}
{\mcitedefaultendpunct}{\mcitedefaultseppunct}\relax
\EndOfBibitem
\bibitem[Rassolov \latin{et~al.}(1998)Rassolov, Pople, Ratner, and
  Windus]{631Gs_Ti_rassolov1998a}
Rassolov,~V.~A.; Pople,~J.~A.; Ratner,~M.~A.; Windus,~T.~L. 6-31G* Basis Set
  for Atoms K through Zn. \emph{J. Chem. Phys.} \textbf{1998}, \emph{109},
  1223--1229\relax
\mciteBstWouldAddEndPuncttrue
\mciteSetBstMidEndSepPunct{\mcitedefaultmidpunct}
{\mcitedefaultendpunct}{\mcitedefaultseppunct}\relax
\EndOfBibitem
\bibitem[Rassolov \latin{et~al.}(2001)Rassolov, Ratner, Pople, Redfern, and
  Curtiss]{631Gs_Ga_rassolov2001a}
Rassolov,~V.~A.; Ratner,~M.~A.; Pople,~J.~A.; Redfern,~P.~C.; Curtiss,~L.~A.
  6-31G* Basis Set for Third-Row Atoms. \emph{J. Comput. Chem.} \textbf{2001},
  \emph{22}, 976--984\relax
\mciteBstWouldAddEndPuncttrue
\mciteSetBstMidEndSepPunct{\mcitedefaultmidpunct}
{\mcitedefaultendpunct}{\mcitedefaultseppunct}\relax
\EndOfBibitem
\bibitem[Eichkorn \latin{et~al.}(1997)Eichkorn, Weigend, Treutler, and
  Ahlrichs]{weigend_aux_basis}
Eichkorn,~K.; Weigend,~F.; Treutler,~O.; Ahlrichs,~R. Auxiliary Basis Sets for
  Main Row Atoms and Transition Metals and Their Use to Approximate Coulomb
  Potentials. \emph{Theor. Chem. Acc} \textbf{1997}, \emph{97}, 119\relax
\mciteBstWouldAddEndPuncttrue
\mciteSetBstMidEndSepPunct{\mcitedefaultmidpunct}
{\mcitedefaultendpunct}{\mcitedefaultseppunct}\relax
\EndOfBibitem
\bibitem[Lehtola(2019)]{lehtola_cholesky_orth_jcp_2019}
Lehtola,~S. {Curing Basis Set Overcompleteness with Pivoted Cholesky
  Decompositions}. \emph{{J}. {C}hem. {P}hys.} \textbf{2019}, \emph{151},
  241102\relax
\mciteBstWouldAddEndPuncttrue
\mciteSetBstMidEndSepPunct{\mcitedefaultmidpunct}
{\mcitedefaultendpunct}{\mcitedefaultseppunct}\relax
\EndOfBibitem
\bibitem[Pipek and Mezey(1989)Pipek, and Mezey]{pipek_mezey_jcp_1989}
Pipek,~J.; Mezey,~P.~G. {A Fast Intrinsic Localization Procedure Applicable for
  {\it{Ab Initio}} and Semiempirical Linear Combination of Atomic Orbital Wave
  Functions}. \emph{{J}. {C}hem. {P}hys.} \textbf{1989}, \emph{90}, 4916\relax
\mciteBstWouldAddEndPuncttrue
\mciteSetBstMidEndSepPunct{\mcitedefaultmidpunct}
{\mcitedefaultendpunct}{\mcitedefaultseppunct}\relax
\EndOfBibitem
\bibitem[Not()]{Note-1}
In constructing IAOs, the MINAO basis was used for all elements but Ba, for
  which the ANO-RCC-MB minimal basis was used instead.\relax
\mciteBstWouldAddEndPunctfalse
\mciteSetBstMidEndSepPunct{\mcitedefaultmidpunct}
{}{\mcitedefaultseppunct}\relax
\EndOfBibitem
\bibitem[Eriksen()]{decodense}
Eriksen,~J.~J. {{\texttt{decodense}}: A Decomposed Mean-Field Theory Code.
  {\url{https://github.com/januseriksen/decodense}}}\relax
\mciteBstWouldAddEndPuncttrue
\mciteSetBstMidEndSepPunct{\mcitedefaultmidpunct}
{\mcitedefaultendpunct}{\mcitedefaultseppunct}\relax
\EndOfBibitem
\bibitem[Willand \latin{et~al.}(2013)Willand, Kvashnin, Genovese,
  V{\'a}zquez-Mayagoitia, Deb, Sadeghi, Deutsch, and
  Goedecker]{coh_en_PP_goedecker}
Willand,~A.; Kvashnin,~Y.~O.; Genovese,~L.; V{\'a}zquez-Mayagoitia,~{\'A}.;
  Deb,~A.~K.; Sadeghi,~A.; Deutsch,~T.; Goedecker,~S. {Norm-Conserving
  Pseudopotentials with Chemical Accuracy Compared to All-Electron
  Calculations}. \emph{J. Chem. Phys.} \textbf{2013}, \emph{138}, 104109\relax
\mciteBstWouldAddEndPuncttrue
\mciteSetBstMidEndSepPunct{\mcitedefaultmidpunct}
{\mcitedefaultendpunct}{\mcitedefaultseppunct}\relax
\EndOfBibitem
\bibitem[Strite and Morko{\c{c}}(1992)Strite, and Morko{\c{c}}]{GaN_review}
Strite,~S.; Morko{\c{c}},~H. GaN, AlN, and InN: A Review. \emph{J. Vac. Sci.
  Technol. B.} \textbf{1992}, \emph{10}, 1237\relax
\mciteBstWouldAddEndPuncttrue
\mciteSetBstMidEndSepPunct{\mcitedefaultmidpunct}
{\mcitedefaultendpunct}{\mcitedefaultseppunct}\relax
\EndOfBibitem
\bibitem[Not()]{Note-2}
Structures have been loaded from the Materials Project database under the
  identifiers mp-804, mp-830, mp-1007824, and mp-2853.\relax
\mciteBstWouldAddEndPunctfalse
\mciteSetBstMidEndSepPunct{\mcitedefaultmidpunct}
{}{\mcitedefaultseppunct}\relax
\EndOfBibitem
\bibitem[Not()]{Note-3}
Structures have once again been loaded from the Materials Project database
  under the identifiers mp-5020, mp-5777, and mp-5986.\relax
\mciteBstWouldAddEndPunctfalse
\mciteSetBstMidEndSepPunct{\mcitedefaultmidpunct}
{}{\mcitedefaultseppunct}\relax
\EndOfBibitem
\bibitem[Not()]{Note-4}
We note how our charges differ from those in Ref.
  \citenum{dronskowski_jpcc_2021}, but not by more than what is to be expected
  when considering the conceptual differences between Mulliken and L{\"o}wdin
  procedures for calculating these.\relax
\mciteBstWouldAddEndPunctfalse
\mciteSetBstMidEndSepPunct{\mcitedefaultmidpunct}
{}{\mcitedefaultseppunct}\relax
\EndOfBibitem
\bibitem[J{\'o}nsson \latin{et~al.}(2017)J{\'o}nsson, Lehtola, Puska, and
  J{\'o}nsson]{lehtola_jonsson_pmwf_jctc_2017}
J{\'o}nsson,~E.~{\"O}.; Lehtola,~S.; Puska,~M.; J{\'o}nsson,~H. {Theory and
  Applications of Generalized Pipek–Mezey Wannier Functions}. \emph{{J}.
  {C}hem. {T}heory {C}omput.} \textbf{2017}, \emph{13}, 460\relax
\mciteBstWouldAddEndPuncttrue
\mciteSetBstMidEndSepPunct{\mcitedefaultmidpunct}
{\mcitedefaultendpunct}{\mcitedefaultseppunct}\relax
\EndOfBibitem
\bibitem[Cui \latin{et~al.}(2020)Cui, Zhu, and Chan]{cui_period_dmet_jctc_2020}
Cui,~Z.-H.; Zhu,~T.; Chan,~G. K.-L. {Efficient Implementation of {\textit{Ab
  Initio}} Quantum Embedding in Periodic Systems: Density Matrix Embedding
  Theory}. \emph{{J}. {C}hem. {T}heory {C}omput.} \textbf{2020}, \emph{16},
  119\relax
\mciteBstWouldAddEndPuncttrue
\mciteSetBstMidEndSepPunct{\mcitedefaultmidpunct}
{\mcitedefaultendpunct}{\mcitedefaultseppunct}\relax
\EndOfBibitem
\bibitem[Clement \latin{et~al.}(2021)Clement, Wang, and
  Valeev]{valeev_pmwf_jctc_2021}
Clement,~M.~C.; Wang,~X.; Valeev,~E.~F. {Robust Pipek–Mezey Orbital
  Localization in Periodic Solids}. \emph{{J}. {C}hem. {T}heory {C}omput.}
  \textbf{2021}, \emph{17}, 7406\relax
\mciteBstWouldAddEndPuncttrue
\mciteSetBstMidEndSepPunct{\mcitedefaultmidpunct}
{\mcitedefaultendpunct}{\mcitedefaultseppunct}\relax
\EndOfBibitem
\bibitem[Sch{\"a}fer \latin{et~al.}(2021)Sch{\"a}fer, Gallo, Irmler, Hummel,
  and Gr{\"u}neis]{schaefer_iao_wannier_jcp_2021}
Sch{\"a}fer,~T.; Gallo,~A.; Irmler,~A.; Hummel,~F.; Gr{\"u}neis,~A. {Surface
  Science Using Coupled Cluster Theory via Local Wannier Functions and
  in-RPA-Embedding: The Case of Water on Graphitic Carbon Nitride }. \emph{{J}.
  {C}hem. {P}hys.} \textbf{2021}, \emph{155}, 244103\relax
\mciteBstWouldAddEndPuncttrue
\mciteSetBstMidEndSepPunct{\mcitedefaultmidpunct}
{\mcitedefaultendpunct}{\mcitedefaultseppunct}\relax
\EndOfBibitem
\bibitem[Zhu and Tew(2024)Zhu, and Tew]{zhu_tew_bloch_iao_arxiv_2024}
Zhu,~A.; Tew,~D.~P. {Wannier Function Localisation Using Bloch Intrinsic Atomic
  Orbitals}. 2024; arXiv:2407.00852\relax
\mciteBstWouldAddEndPuncttrue
\mciteSetBstMidEndSepPunct{\mcitedefaultmidpunct}
{\mcitedefaultendpunct}{\mcitedefaultseppunct}\relax
\EndOfBibitem
\end{mcitethebibliography}
\end{document}